\documentclass{JHEP3}

 \usepackage{epsfig}

 \title{
  Hadron pair photoproduction within the Veneziano model
  }

 \author{Kosuke Odagiri\\
  Institute of Physics, Academia Sinica, Nankang, Taipei, Taiwan 11529,
  The Republic of China}

 \abstract{
  We first suppose that low-energy hadron pair photoproduction reactions
$\gamma^{(*)}\gamma^{(*)}\to h\bar h$ are dominated by $s$-channel
resonance contributions.
  Their normalization is then calculated by their correspondence with the
Reggeon term in the Regge parametrization of the $\gamma h$ total cross
sections.
  For the case of $p\bar p$, we make use of the measured $\gamma p$ total 
cross section, and for the case of $K^+K^-$, we make use of the 
corresponding total cross section that is estimated using Regge 
factorization.
  For hadrons that have no such data, we can only provide rough estimation
based on the additive quark rule.
  As an effective approach that is convenient and parameter-free, we adopt
the Veneziano model in the simplest form.
  The model is only applicable to the region of low centre-of-mass energy.  
When the transverse momentum is large, perturbative QCD takes over,
whereas in the Regge region, it is known that the Regge pole picture fails
in photoproduction.
  Despite the shortcomings of the model, we find that the parameter-free
amplitudes offer a sound description of the data at hand.
  }

 \keywords{nef.pmo}

 \preprint{hep-ph/0406267}

\begin{document}

 \section{Introduction}

 \subsection{Experimental status}

  Belle has accumulated large statistics in hadron-pair photoproduction 
event samples, in particular, of $\gamma^{(*)}\gamma^{(*)}\to p\bar p$ 
\cite{bellep} and $K^+K^-$ \cite{bellek}.
  In both cases, the measured cross section and the angular distribution 
are largely in agreement with the previous measurements 
\cite{venus,cleo,tpc,argus}.
  As an example, we show the results\footnote{We thank C.C.~Kuo for their 
kind permission to reproduce their figures in this paper. The discussions 
we present in this paper are, however, based on a more recent analysis of 
ref.~\cite{bellep}.} \cite{photon03} for the $p\bar p$ measured total 
cross section in fig.~\ref{fig_belle_crosssec} and the angular 
distribution in fig.~\ref{fig_belle_angulardis}. We should note that the 
difference between the Belle measurement and the other experiments has 
been resolved and the latest results \cite{bellep} are largely in 
agreement with the data from VENUS \cite{venus} and CLEO \cite{cleo}.

 \FIGURE[ht]{
 \epsfig{file=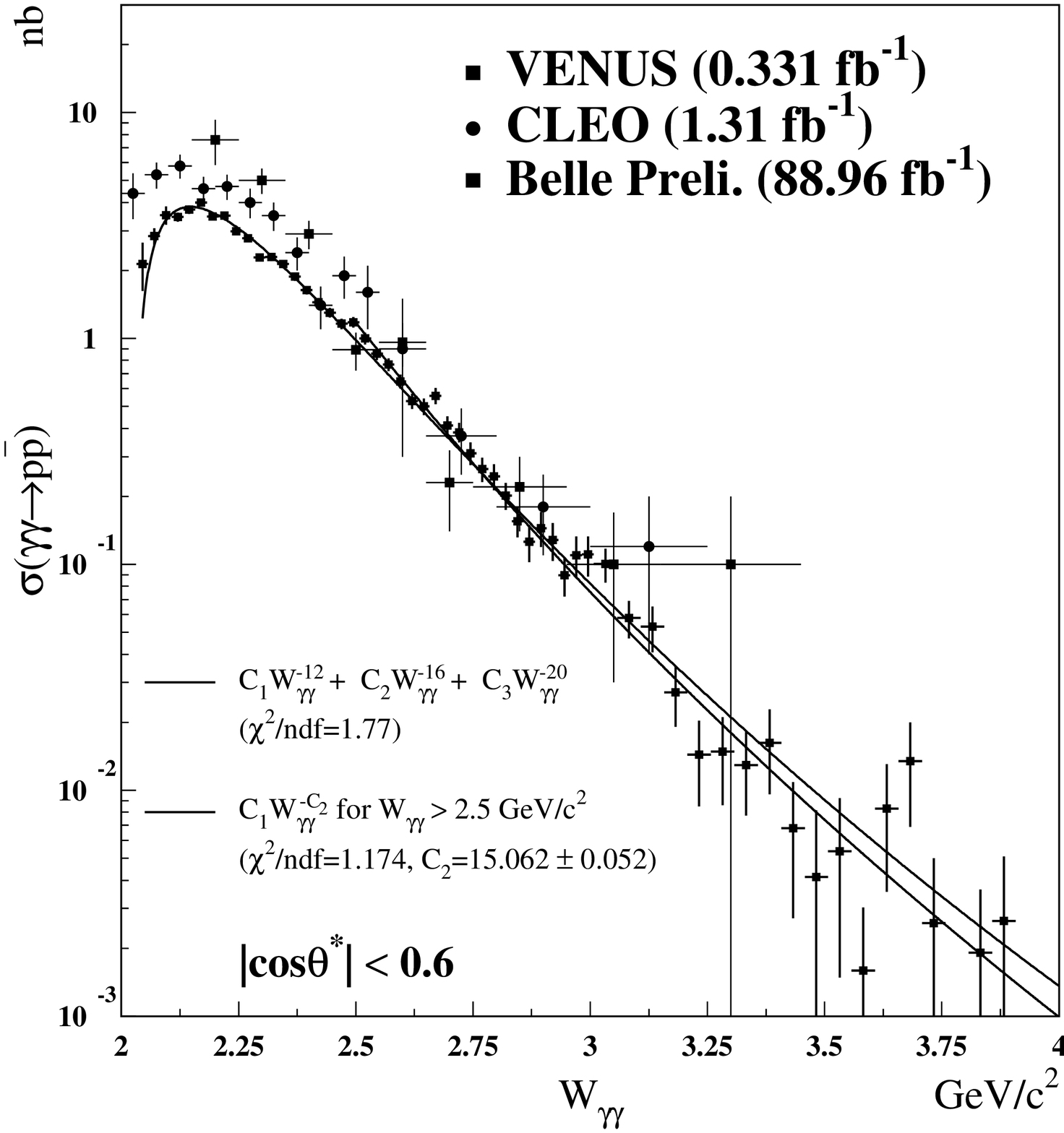,width=10cm}
 \caption{$\gamma\gamma\to p\bar p$ cross section versus centre-of-mass 
energy at VENUS, CLEO and Belle in the central region, defined by 
$|\cos\theta^*|<0.6$. The vertical error-bars on the Belle data are due to 
the statistical error in the event and the Monte Carlo samples only.
 \label{fig_belle_crosssec}}}

 \FIGURE[ht]{
 \epsfig{file=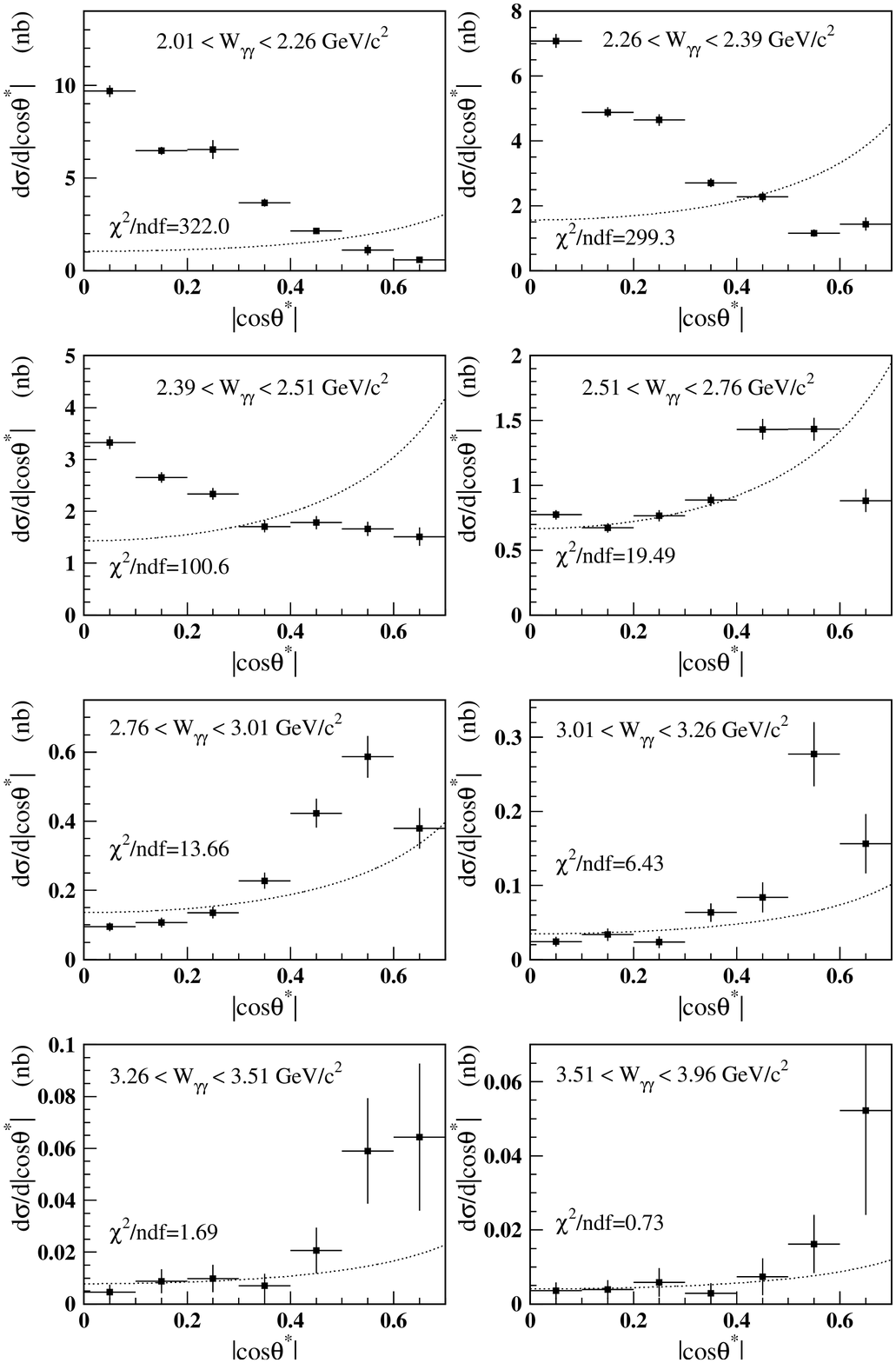,width=11cm}
 \caption{Angular distribution of $\gamma\gamma\to p\bar p$ events at 
Belle. The vertical error-bars represent the statistical uncertainty only.
 \label{fig_belle_angulardis}}}

  The Belle measurements are restricted to the central region defined by
$|\cos\theta^*|<0.6$, where $\theta^*$ is the polar angle in the
centre-of-mass frame of the produced hadron pair.
  The total cross section in this region as a function of the effective
centre-of-mass energy $W_{\gamma\gamma}$ shows prominent resonance
structure for the case of $K^+K^-$, and a near-continuum structure for the
case of $p\bar p$ although some remnants of individual resonances can
still be seen. This is in accord with the expectation from the fact that
the proton is heavier than the kaon and that the mass difference between
resonances is small when $W_{\gamma\gamma}$ is large.

  The angular distributions show a more marked trend. At sufficiently high
$W_{\gamma\gamma}$, the events are concentrated in the forward region as
expected on general grounds from $t$-channel exchange be it of
perturbative or nonperturbative nature, but just above the threshold, the
distributions are peaked in the transverse direction, i.e.,
$\cos\theta^*\sim0$. This latter behaviour would be very difficult to
explain in a perturbative analysis but is easily accommodated in the
resonance picture, as follows.

  Let us consider the simpler case of the reaction $\gamma\gamma\to 
K^+K^-$, and suppose that it proceeds by an $s$-channel spin-2 resonance, 
for example $f_2(1270)$, $a_2(1320)$ or $f'_2(1525)$. As $K^\pm$ are 
spin-0, there are altogether only two independent helicity combinations, 
due to the initial-state photon helicities. The net $s$-channel helicity 
is either $\pm2$ or 0 along the direction of the photons.
  The angular distributions are then proportional to the square of:
 \begin{eqnarray}
  d^2_{2,0} &=& \frac{\sqrt{6}}4\sin^2\theta^*, \\
  d^2_{0,0} &=& \frac32\cos^2\theta^*-\frac12.
 \end{eqnarray}
  For real photons, the two amplitudes do not interfere because of 
different photon helicities. We thus see that the distribution has a local 
maximum at $\cos\theta^*=0$ in both cases. For the latter amplitude, the 
distribution also has maxima at $\cos\theta^*=\pm1$. The $K^+K^-$ cross 
section just above the threshold indeed shows a marked peak corresponding 
to $f'_2(1525)$ and so it is inevitable that the distribution is peaked at 
$\cos\theta^*=0$.

  This above argument provides good reason to suppose that the process is
predominantly determined by resonance dynamics in the energy range probed
by Belle. However, the $p\bar p$ cross section seems to have a $\sim
W_{\gamma\gamma}^{-10}$ behaviour expected from the quark-counting rule
and so, at first sight, seems to have a perturbative description too.

  The quark-counting rule states that at high energy, the large-angle
$2\to2$ scattering cross section is described by:
 \begin{equation}
  \frac{d\sigma}{dt}\Big|_\mathrm{\cos\theta^*=const.}
  =\frac{f(\cos\theta^*)}{s^{K-2}},
  \label{eqn_quark_counting}
 \end{equation}
  where $K$ is the number of `elementary fields', i.e., in our case, 
quarks and photons, in the external legs. For meson pair production, we 
have $K=6$, and for baryon pair production, we have $K=8$. $s$ in this 
case is $W_{\gamma\gamma}^2$.
  Hence the cross section, integrated over $|\cos\theta^*|<0.6$, should 
scale as $W_{\gamma\gamma}^{-10}$ for baryon pair photoproduction and 
$W_{\gamma\gamma}^{-6}$ for meson pair photoproduction.

  There are two problems with this application to the case of $p\bar p$. 
  First, contrary to eqn.~(\ref{eqn_quark_counting}), the angular 
distribution depends on $W_{\gamma\gamma}$. In fact, it is found 
\cite{venus,l3ppbar,kakusan} that in the more central region 
$|\cos\theta|<0.3$, the fall-off is more rapid whereas one would expect 
the quark-counting rule to work better in this region.
  Second, fitting the data above 2.5 GeV with the power kept as a free
parameter yields $W_{\gamma\gamma}^{-14.87}$ \cite{bellep} rather than
$W_{\gamma\gamma}^{-10}$. It seems that at 2.5 GeV, at least, the
quark-counting rule is not strictly applicable.

 \subsection{An approach based on the Veneziano model}

  In the light of the above, it seems natural to consider an approach
based on resonance dynamics borrowing some ideas from Regge theory
\cite{collins}. For the case of $p\bar p$, if the $f/a$ mesons are
responsible, their contribution can be normalized through the optical
theorem by finding the contribution of the $f/a$ trajectories to the
$\gamma p$ total cross section which is given by \cite{ddln,dlcs}:
 \begin{equation}
  \sigma_\mathrm{tot}(\gamma p)=0.0677s^{0.0808}+0.129s^{-0.4525}.
  \label{eqn_gamptot}
 \end{equation}
  The cross section is measured in mb and $s$ is measured in GeV$^2$. The 
former term is the soft pomeron contribution and the latter term is the 
contribution of $f_2$ and $a_2$ trajectories, of which we know that $f_2$ 
is the dominant \cite{ddln} so that we expect the $\gamma n$ total cross 
section to be almost identical.

  As an economical way of writing the many resonance contributions, we
propose to adopt the Veneziano model \cite{veneziano}.

  One advantage in adopting the Veneziano model is that the amplitude is
dual by construction, and so, in principle, the transition from the
$s$-channel resonance picture to the $t$-channel pole picture should be
consistent and suffer from no double-counting as would be the case had we 
simply summed the $s$-channel and $t$-channel contributions together.

  Having said that, there are two pitfalls that one should bear in mind
throughout.

   First, the Veneziano model is constructed for spin-less meson scattering
processes. The model would perform badly where the helicities of the
particles matter. For example, for baryon exchange processes, there is
incorrect behaviour in the Regge limit as compared with the Regge pole
expectations.

  Second, photoproduction processes, such as $\gamma p\to \pi^0 p$, are 
known to violate the simple Regge pole picture. Instead of the usual 
$s^{\alpha(t)}$ behaviour of the amplitude in the Regge limit, i.e., large 
$s$ and small fixed $t$, it has been found that $d\sigma/dt$ goes as 
$s^{-2}$ \cite{ddln} when meson exchange is expected and $s^{-3}$ 
\cite{storrow} when baryon exchange is expected. In both cases, no sign of 
`shrinkage' is found, i.e., the amplitude, as a function of $t$, goes not 
as $s^{\alpha(t)}$ but as $C^{\alpha(t)}$ where $C$ is constant.
  It is worth noting here that the other baryon exchange processes are 
also known to share these same characteristics \cite{storrow}.

  We hence see that our picture should work only at low $W_{\gamma\gamma}$
where the perturbative picture is still not applicable and, at the same
time, we are safely away from the Regge region where the Regge pole
picture presumably breaks down. This last point, however, merits further
experimental investigation.

  In addition, even when $W_{\gamma\gamma}$ is small, there would be 
problems when the helicity structure is important. One other point to bear 
in mind is that the resonances do not lie exactly on linear trajectories 
and so there would be some error coming from inaccurate parametrization. 
We should emphasize here that what we are dealing with is just a simple 
`toy model' of resonance-and-pole-dominated physics, and in any case the 
economy of the model is sufficiently attractive to merit the analysis of 
its instigation.

  We perform several simple calculations to show that despite all these
restrictions, the result shows striking resemblance to real data, though
the quantitative description leaves room for improvement.

  The paper is organized as follows. We write down the amplitude in 
sec.~\ref{sec_veneziano}. The result of calculations is shown in 
sec.~\ref{sec_result}, together with the discussion of our findings. The 
conclusions are stated at the end.

 \section{The Veneziano amplitude}\label{sec_veneziano}

  The Veneziano amplitude \cite{veneziano} is an amplitude with the pole 
structure inspired by Regge theory and conforms to the idea of duality 
\cite{collins}. For meson trajectories, it reproduces the Regge behaviour 
at fixed $t$ in the limit of large $s$. For baryon trajectories this does 
not work and, if we desire to recover Regge behaviour we must, for 
example, insert an extra $1/\sqrt{s}$ factor in the amplitude.

  However, it is known that processes involving baryon exchange are not
described well by the Regge pole picture \cite{storrow}. Furthermore,
photoproduction processes are not described well by the Regge pole picture
even when mesons are exchanged, as in the process $\gamma p \to\pi^0 p$
\cite{ddln}.

 \subsection{The formalism}

  Let us write down our amplitudes as follows. Meson pair photoproduction
is given by:
 \begin{equation}
  A(\gamma\gamma\to M\overline M)=\frac{\overline\beta}\pi
  \frac{\Gamma(1-\alpha_s(s))\Gamma(1-\alpha_t(t))}
  {\Gamma(1-\alpha_s(s)-\alpha_t(t))} + (t\leftrightarrow u).
  \label{eqn_mesonamp}
 \end{equation}
  Baryon pair photoproduction is given by:
 \begin{equation}
  A(\gamma\gamma\to B\overline B)=\frac{\overline\beta}\pi
  \frac{\Gamma(1-\alpha_s(s))\Gamma(\frac12-\alpha_t(t))}
  {\Gamma(\frac12-\alpha_s(s)-\alpha_t(t))} + (t\leftrightarrow u).
  \label{eqn_baryonamp}
 \end{equation}
  $\Gamma$ is the Euler Gamma function.
$\alpha_s(s)=\alpha_s(0)+\alpha_s's$ is the $s$ channel trajectory (and
certainly should not be confused with the strong coupling $\alpha_S$ to
which it has no direct relation). $\alpha_t(t)=\alpha_t(0)+\alpha_t't$ is
the $t$ channel trajectory. Both trajectories need to be linear.
  $\overline\beta$ is the overall coupling, which we may assume to be
constant for simplicity.

  In order to obtain the leading contribution, one should set
$\alpha_t(t)$ to be the leading trajectory, i.e., the trajectory with the
greatest intercept $\alpha_t(0)$, consistent with the $t$-channel quantum
numbers.
  For the $s$-channel, one should include all trajectories that have
significant contribution. We have assumed in the above that the lightest
member of the $s$-channel trajectory is spin-1 and the lightest member of
the $t$-channel trajectory is spin-1 or spin-1/2.
  However, the leading $s$-channel contribution is spin-2 by symmetry
after summing the $t$-channel and the $u$-channel contributions together,
so that there is no contribution from, for example, $\rho$ and $\omega$.

 \subsection{Remarks on the amplitude}

  We note that the $s\leftrightarrow t$ crossed amplitude corresponding to 
eqn.~(\ref{eqn_baryonamp}) does not have correct angular behaviour 
expected from fermionic resonances \cite{storrow}. This is the reason for 
the incorrect Regge behaviour in the uncrossed amplitude.

  The Regge limit, $s\to\infty,t$ or $u=$const., for the above amplitudes 
is obtained using the Stirling formula in the form:
 \begin{equation}
  \Gamma(x+a)/\Gamma(x+b)=x^{a-b}\left(1+\mathcal{O}(x^{-1})\right).
  \label{eqn_stirling}
 \end{equation}
  For the two amplitudes, we obtain respectively:
 \begin{eqnarray}
  A(\gamma\gamma\to M\overline M)&\approx&\frac{\overline\beta}\pi
  \Gamma\left(1-\alpha_t(t)\right)
  \left(-\alpha_s(s)\right)^{\alpha_t(t)}
  + (t\leftrightarrow u), \label{eqn_mesonic_regge}\\
  A(\gamma\gamma\to B\overline B)&\approx&\frac{\overline\beta}\pi
  \Gamma\left(\frac12-\alpha_t(t)\right)
  (-\alpha_s(s))^{1/2+\alpha_t(t)}
  + (t\leftrightarrow u). \label{eqn_baryonic_regge}
 \end{eqnarray}
  Only the former has the behaviour expected from Regge theory. The latter 
has an extra $1/2$ in the power.

  The mathematical properties of the Veneziano amplitude are well 
documented elsewhere \cite{collins,ddln,veneziano}, but it is worth 
restating the manifestation of duality in it.
  The Veneziano amplitude is dual in the sense that we may expand it 
either in terms of the $s$-channel resonances or the $t$-channel poles, 
and when expanded in terms of the $s$-channel resonances, no $t$-channel 
poles appear, and vice versa.
  For the Euler beta function $B(x,y)$, following the notation of 
ref.~\cite{ddln}, we have:
 \begin{equation}
  B(x,y)\equiv
  \frac{\Gamma(x)\Gamma(y)}{\Gamma(x+y)}\equiv
  \sum_{n=0}^\infty
  \frac{(-1)^n}{n!}\frac{1}{x+n}\frac{\Gamma(y)}{\Gamma(y-n)}\equiv
  \sum_{n'=0}^\infty
  \frac{(-1)^{n'}}{{n'}!}\frac{1}{y+n'}\frac{\Gamma(x)}{\Gamma(x-n')}.
  \label{eqn_bxy}
 \end{equation}
  We see that when expanded in terms of poles in $x$, no poles in $y$
appear, and vice versa. The amplitudes (\ref{eqn_mesonamp}),
(\ref{eqn_baryonamp}) are related trivially to $B(x,y)$ and so no double
poles appear here also. We express this diagrammatically, for the baryonic
case, in fig.~\ref{fig_duality}.

 \FIGURE[ht]{
 \epsfig{file=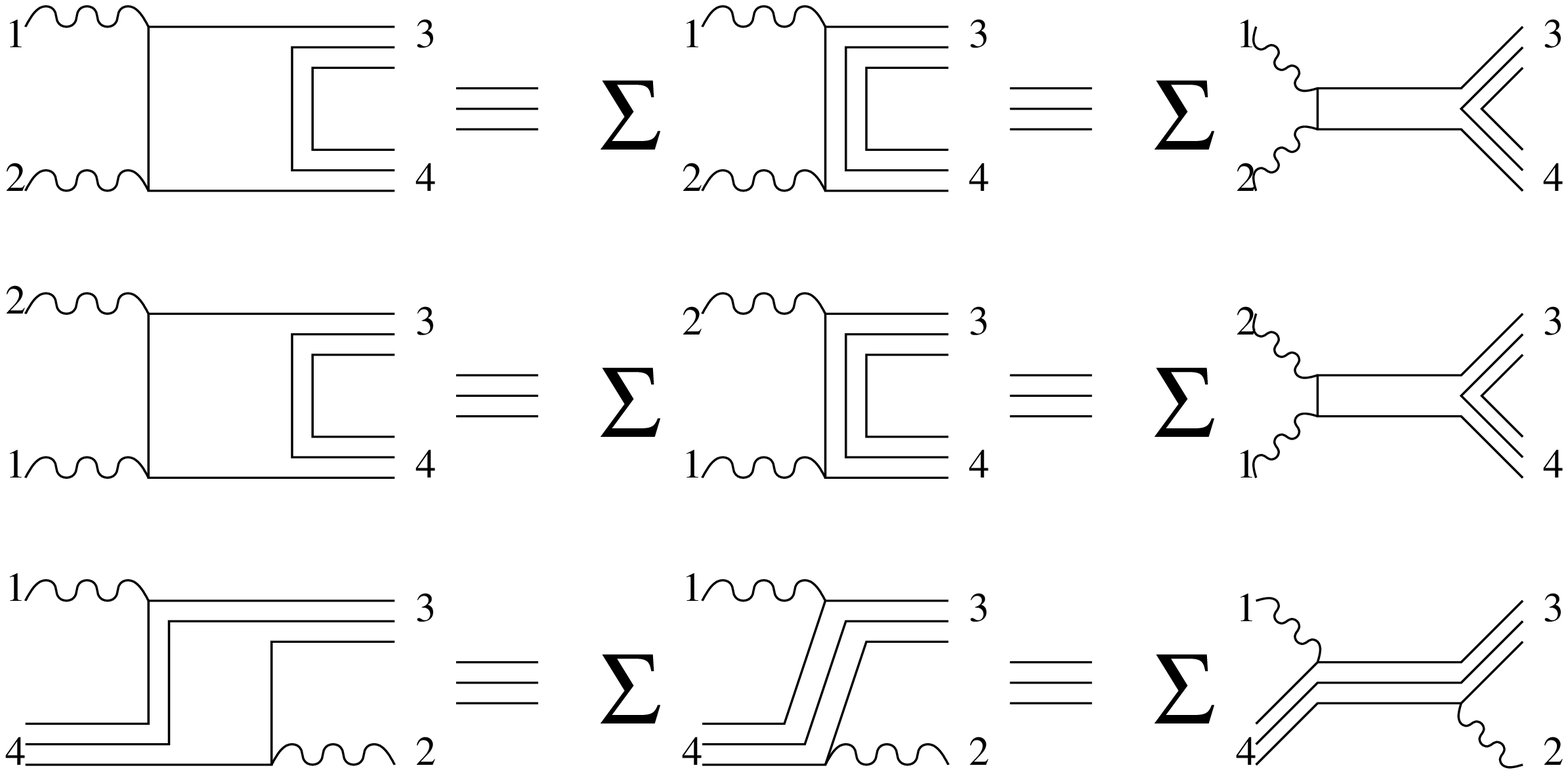,width=13cm}
 \caption{The quark-line diagrams for the three dual amplitudes in
$\gamma\gamma\to p\bar p$. The diagrams shown are for the $s-t$ amplitude
(top), the $s-u$ amplitude (middle) and the $t-u$ amplitude (bottom).
 \label{fig_duality}}}

  In fig.~\ref{fig_duality}, in addition to the terms included in 
eqns.~(\ref{eqn_mesonamp}) and (\ref{eqn_baryonamp}), a third class of 
terms is shown, which are $t-u$ dual. We do not consider this class of 
terms, because we wish to consider only terms that have a $s$-channel 
description and because we know that the Regge $t$-channel pole picture 
fails in photoproduction.
  For the same reason, we should emphasize that although the amplitudes as 
written above are expected to entail a smooth transition from a low-energy 
$s$-channel dominated picture to the Regge $t$-channel pole picture, one 
should regard this only as an indication of the enhancement in the forward 
region and not as a prediction of how the forward amplitude should rise.

  We also note that the pole expansion in eqn.~(\ref{eqn_bxy}) requires
the existence of daughter trajectories that contain hadrons with
approximately the same masses as the hadrons on the parent trajectory and
lower $J$.

 \subsection{The trajectories}

  The trajectories that we consider are as follows. For the $s$-channel 
$a/f$ trajectories, we adopt:
 \begin{equation}
  \alpha_{a/f}(s)=0.5+0.4i+0.9s.        \label{eqn_traj_af}
 \end{equation}
  For the real part of the Reggeon slope and intercept, we used the values 
suggested in ref.~\cite{ddln} that applies to the case when 
$\rho/\omega/a/f$ are all on a universal trajectory. Without this 
constraint, ref.~\cite{ddln} suggests $0.70+0.80s$ for the $a/f$ 
trajectories. The imaginary part has been chosen so as to yield a width of 
approximately 200 MeV for resonances at about 2 GeV.
  $s$ is measured in GeV$^2$ as mentioned before.
  As resonances occur for positive integer values of $\alpha(s)$, the 
resonances are at:
 \begin{equation}
  M_{a/f}=\sqrt{(n-0.5)/0.9}\ \mathrm{GeV},\quad (n=1,2,3,\ldots).
  \label{eqn_masses}
 \end{equation}
  We note that although the resonances from the leading trajectory are 
only manifest at even integer values of $n$, there are resonances for 
every value of $n$, on the daughter trajectories. We tabulate the first 
few resonance masses in tab.~\ref{tab_f_masses}

 \TABLE{
  \begin{tabular}{c|ccccc}
  \hline
  n         & 1     & 2           & 3     & 4           & 5 \\
  \hline
  $M$ / GeV & 0.745 & 1.291       & 1.667 & 1.972       & 2.236 \\
  PDG       &       & $f_2(1270)$ &       & $f_4(2050)$ &       \\
  \hline
  \end{tabular}
  \label{tab_f_masses}
  \caption{The first few resonance masses on the trajectory $0.5+0.9s$, 
and the known leading resonances from ref.~\cite{pdg}. The masses of 
$f_2(1270)$ and $f_4(2050)$ are $1275.4\pm1.2$ MeV and $2025\pm8$ MeV, 
respectively. The assignment to the daughter trajectories can be found in 
ref.~\cite{ddln}.}
  }

 \FIGURE[ht]{
 \epsfig{file=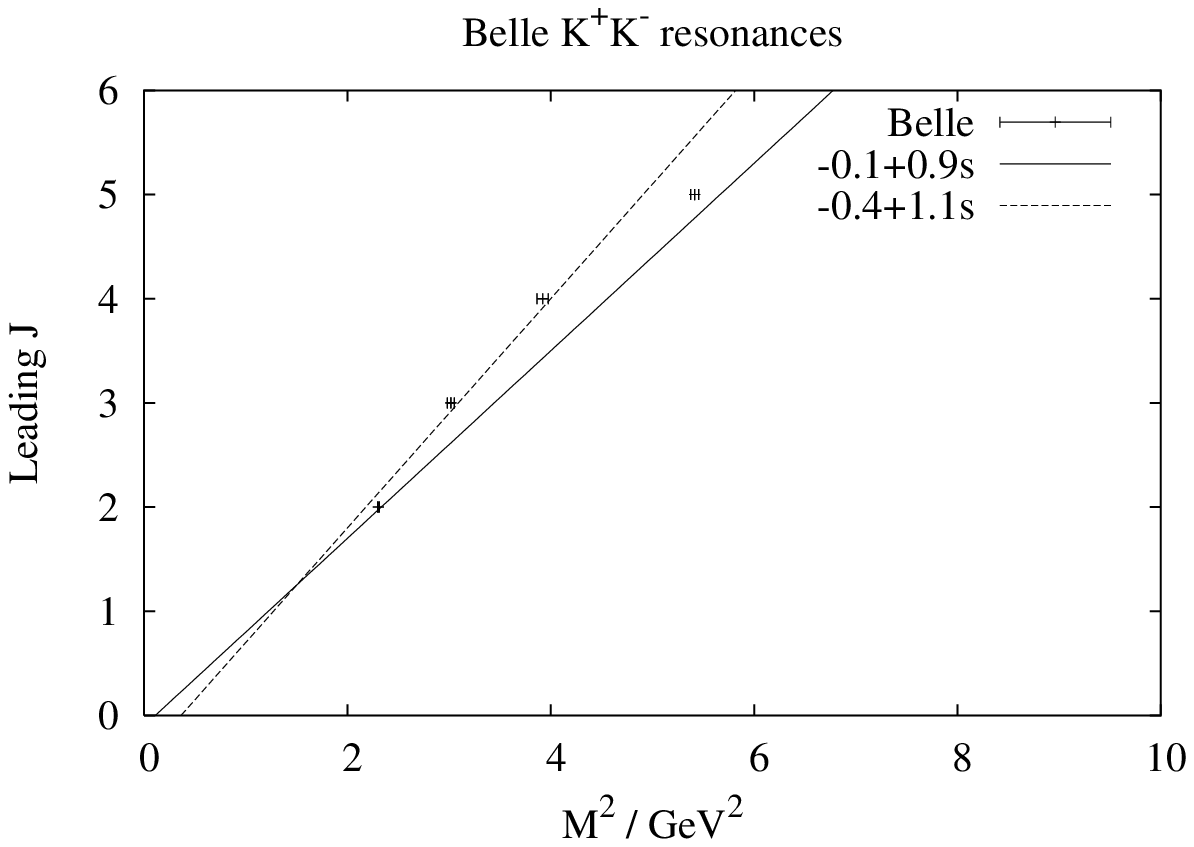,width=9.5cm}
 \epsfig{file=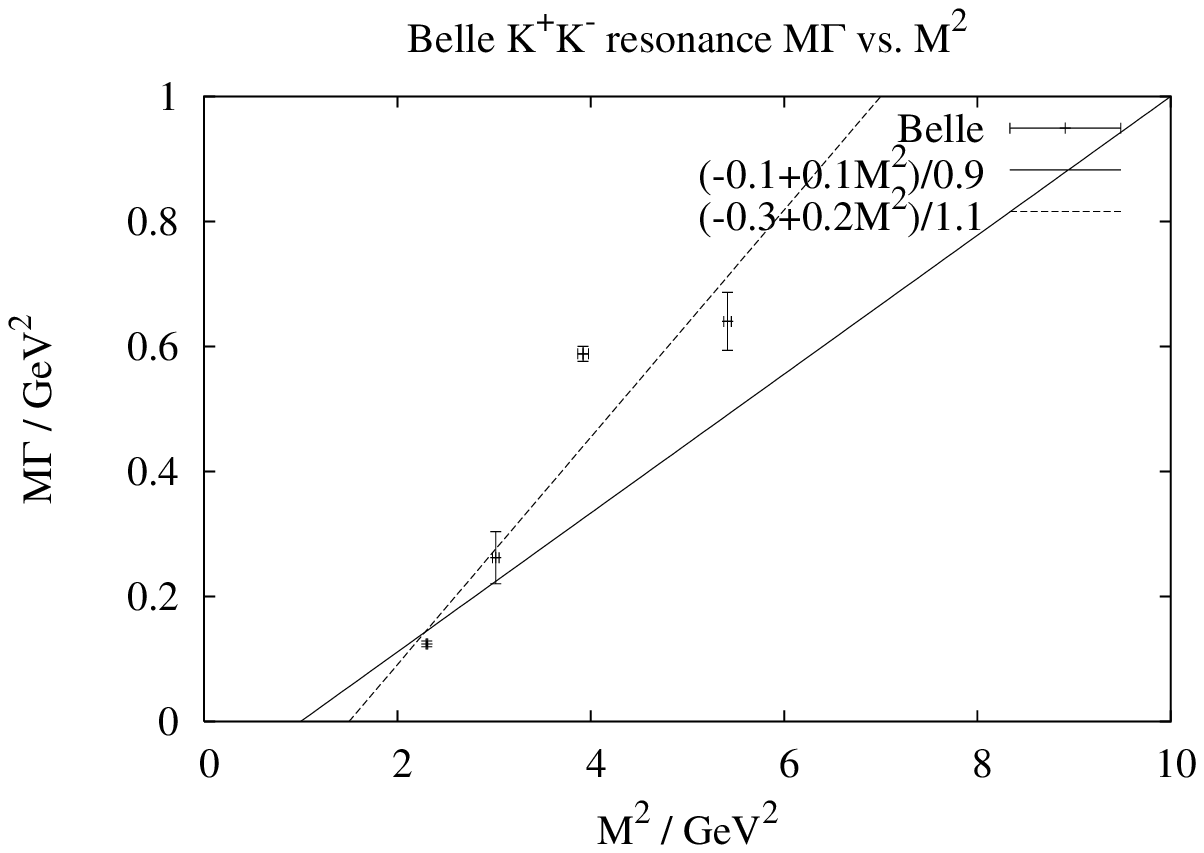,width=10cm}
 \caption{Resonances in $\gamma\gamma\to K^+K^-$ found in 
ref.~\cite{bellek}. The first plot (above) shows a `guess' for the leading 
trajectory and its parametrizations, if these resonances belong to a 
family. The second plot (below) shows the resonance widths against our 
parametrizations. The solid line in each case is the parametrization 
adopted for simulation.
 \label{fig_fprime}}}

  In addition, for processes that involve strange quarks, we adopt:
 \begin{equation}
  \alpha_{f'}(s)=-0.1-0.1i+(0.9+0.1i)s. \label{eqn_traj_fprime}
 \end{equation}
  The parametrization is demonstrated in fig.~\ref{fig_fprime}, together 
with another possibility corresponding to 
$\alpha_{f'}(s)=-0.4-0.3i+(1.1+0.2i)s$.
  The values for $J$ shown in fig.~\ref{fig_fprime} are merely guesses for 
the leading members in the trajectory, and not a result of experimental 
analysis. The odd-valued $J$ would correspond to the $\phi$ mesons on the 
parent trajectory.

  The real part was chosen to go through $f'_2(1525)$ with the constraint 
that the slope is 0.9. This roughly follows the resonances found in 
ref.~\cite{bellek}, but it is not the best choice, as can be seen in 
fig.~\ref{fig_fprime}.

  The widths also approximate the values found in ref.~\cite{bellek}, but 
this is again not the best choice. Contrary to eqn.~(\ref{eqn_traj_af}), 
the imaginary part in this case has a slope. This is necessary since 
$f'_2(1525)$ has a relatively small width of about 80 MeV whereas the 
heavier resonances are expected to have greater widths.

  We should bear in mind that the analysis of ref.~\cite{bellek} does not 
fully take into account the possibility that there are further 
contributions which have peak positions near these resonances, such as 
there is in our present analysis from $a$ and $f$ mesons. Hence the masses 
and widths measured therein are not conclusive. This is the reason why we 
chose to put less weight on tuning with the resonances of 
ref.~\cite{bellek} and used a parametrization which, in the author's 
opinion, is more conservative than the other choice shown in 
fig.~\ref{fig_fprime}.

 \TABLE{
  \begin{tabular}{c|ccccc}
  \hline
  n         & 1           & 2           & 3             & 4     & 5 \\
  \hline
  $M$ / GeV & 1.106       & 1.528       & 1.856         & 2.134 & 2.380 \\
  Belle     &             & 1.518       & 1.737         & 1.980 & 2.327 \\
  PDG       & $\phi(1020)$& $f_2'(1525)$& $\phi_3(1850)$&       &       \\
  \hline
  \end{tabular}
  \label{tab_fprime_masses}
  \caption{The first few resonance masses on the trajectory $-0.1+0.9s$, 
compared with the result of the Belle analysis and with some of the 
$\phi/f'$ resonances from ref.~\cite{pdg}.}
  }
  For the sake of comparison, the first few resonance masses according to 
the parametrization of eqn.~(\ref{eqn_traj_fprime}) are tabulated in 
tab.~\ref{tab_fprime_masses} and are compared against the Belle analysis 
and some of the $\phi/f'$ resonances from PDG \cite{pdg}. There is also a 
$\phi(1680)$ in PDG which is difficult to assign. On the other hand, the 
masses of $f_2'(1525)$ and $\phi_3(1850)$ are in agreement with our 
parametrization.

  For the $t$-channel trajectories, we have for the leading $\Delta/N$ 
trajectory:
 \begin{equation}
  \alpha_{\Delta/N}(t)=0.0+0.9t,  \label{eqn_traj_deln}
 \end{equation}
  both for the neutral and charged baryons. This is taken from
ref.~\cite{storrow}.

  For $K^\pm$, the leading trajectory is that containing $K^*(892)$ rather
than $K^\pm$ whose mass is $493.677\pm0.016$ MeV, and we write this as:
 \begin{equation}
  \alpha_{K^*}(t)=0.2+0.9t.          \label{eqn_traj_kstar}
 \end{equation}
  The numbers were fixed here by assuming that the slope is 0.9, the same
as in eqn.~(\ref{eqn_traj_af}). Having said that, this is in good 
agreement with the known $K\*$ mesons, as shown in fig.~\ref{fig_kfamily}.

 \FIGURE[ht]{
 \epsfig{file=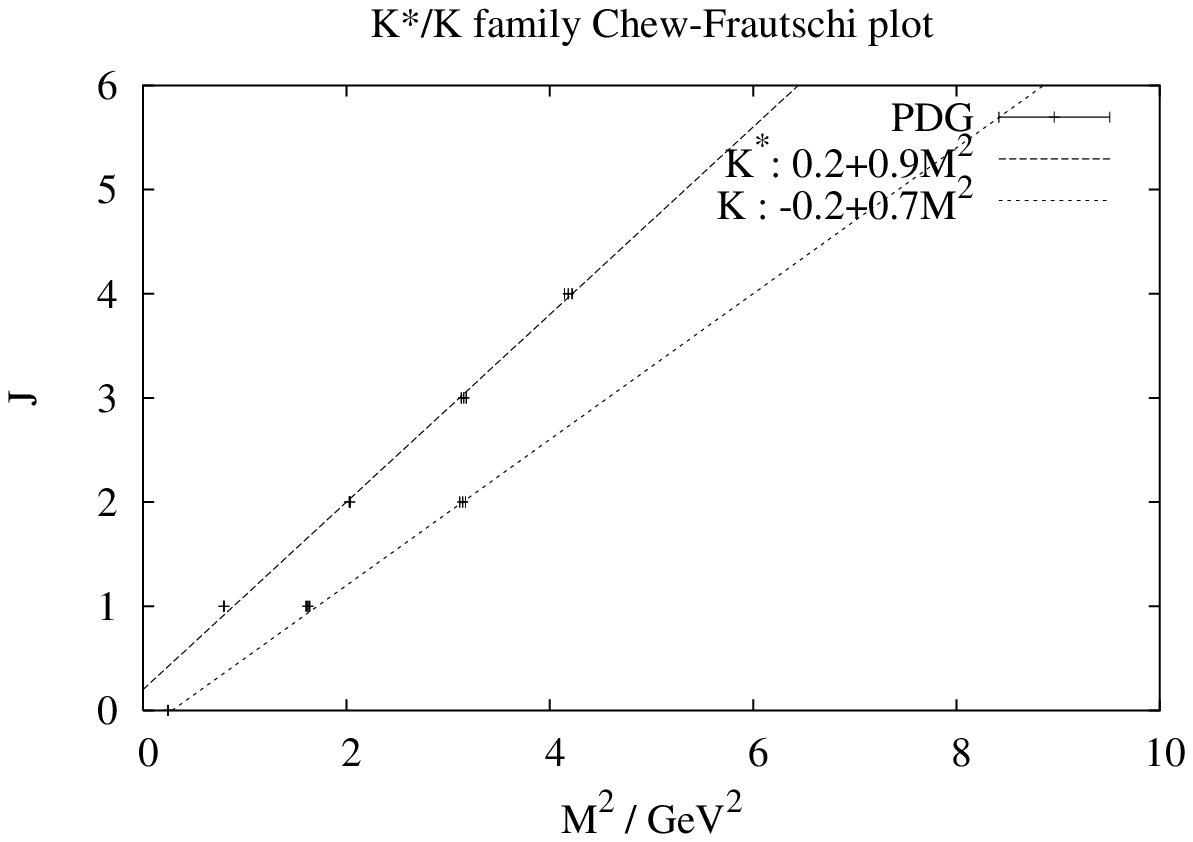,width=12cm}
 \caption{Leading $K$ and $K^*$ trajectories.
 \label{fig_kfamily}}}

  For $\Lambda$ and $\Sigma$ baryons, we have the leading $\Sigma$ 
trajectory given by:
 \begin{equation}
  \alpha_{\Sigma}(t)=-0.27+0.9t.  \label{eqn_traj_sigma}
 \end{equation}
  This is again taken from ref.~\cite{storrow}.

 \subsection{The normalization}\label{sec_normalization}

  Let us now find the normalization factor $\overline\beta$ by considering
the Reggeon mediated part of the photon-hadron total cross sections. The
total cross sections are related to the imaginary parts of the elastic
scattering amplitudes by the optical theorem:
 \begin{equation}
  \sigma_\mathrm{tot}(\gamma h)
  = \frac1s\mathrm{Im}\left[A(\gamma h\to\gamma h)\right]\Big|_{t=0}.
  \label{eqn_optical}
 \end{equation}
  We have not explicitly introduced the spin indices but the averaging
over the amplitudes is implicit. In this case, the amplitudes should be
understood to be the average of the amplitudes in which the $\gamma$ and
$h$ helicities are separately conserved. The helicity-flipping amplitudes
can not be estimated using our present method.

  After crossing the amplitudes (\ref{eqn_mesonamp}),
(\ref{eqn_baryonamp}) to obtain the relevant $\gamma h\to\gamma h$
amplitudes, we take the Regge limit of fixed $t$, in this case $t=0$, and
large $s$ and make use of eqn.~(\ref{eqn_stirling}).
  Both eqns.~(\ref{eqn_mesonamp}) and (\ref{eqn_baryonamp}) have the 
correct Regge behaviour in this limit, and we obtain:
 \begin{equation}
  \sigma_\mathrm{tot}(\gamma h)\approx\frac{\overline\beta}{\pi s}
  \Gamma\left[1-\alpha_s(0)\right]\left(\alpha_t's\right)^{\alpha_s(0)}
  \sin\left[{\pi\alpha_s(0)}\right].
  \label{eqn_ven_totcs}
 \end{equation}
  We note that there is in reality an additional contribution from the 
pomeron which dominates in the large $\sqrt{s}$ limit. Here we limit 
ourselves to the discussion of the $a/f$ meson contributions only. From 
the real part of eqn.~(\ref{eqn_traj_af}), we have the intercept 
$\alpha_s(0)=0.5$. For $\alpha_t'=0.9$ and converting from GeV$^2$ to mb, 
we obtain:
 \begin{equation}
  \sigma_{\mathrm{tot,\ }a/f}(\gamma h)
  \approx 0.208\overline\beta s^{-0.5}\ \mathrm{mb}.
  \label{eqn_ven_totcs_num}
 \end{equation}
  For $\gamma\gamma\to p\bar p$, $\overline\beta$ can then be determined 
by comparing against the second term of eqn.~(\ref{eqn_gamptot}), namely 
$0.129s^{-0.4525}$ mb. However, the powers are slightly different due to 
the different values of the $a/f$ intercept adopted. We choose to equate 
the two at 10 GeV and obtain:
 \begin{equation}
  \overline\beta(\gamma\gamma\to p\bar p)=0.770\ .
 \end{equation}
  We note that the $n\bar n$ process should be almost identical to the
$p\bar p$ case. The $pn$ and $\bar pn$ total cross sections are similar to
the $pp$ and $\bar pp$ total cross sections, indicating that the isospin-1
contribution from the $a$ trajectory is small.

  The differential cross section is given generally by:
 \begin{equation}
  \frac{d\sigma}{d\cos\theta^*}=
  \frac{\sqrt{1-4m_h^2/s}}{32\pi s}\frac{1+2J_h}2
  \frac1{1+\delta}\left|A\right|^2.
 \end{equation}
  $m_h$ is the hadron mass and $J_h$ is the spin. $\delta$ is 1 for
identical particle final states, for example $\gamma\gamma\to\pi^0\pi^0$,
and 0 otherwise.

  We have again assumed that only the helicity conserving amplitudes 
contribute in the sense that, for example, the initial-state photons are 
in the left--right and the right--left combinations, as the other 
amplitudes can not be estimated.

  $t$ and $u$ are calculated from $s$ and $\cos\theta^*$ as usual by:
 \begin{equation}
  t=m_p^2-\frac s2\left(1-\cos\theta^*\sqrt{1-4m_p^2/s}\right),
  \quad
  u=m_p^2-\frac s2\left(1+\cos\theta^*\sqrt{1-4m_p^2/s}\right).
 \end{equation}

  Now let us turn our attention to the case of $\gamma\gamma\to K^+K^-$. 
There is no data for the $\gamma K^\pm$ total cross section, so we must 
estimate it using Regge factorization using the $p K^\pm$ scattering total 
cross section, which goes as \cite{ddln,dlcs}:
 \begin{eqnarray}
  \sigma_\mathrm{tot}(K^+p) &=& 11.93s^{0.0808}+7.58s^{-0.4525}
  \ \mathrm{mb}, \\
  \sigma_\mathrm{tot}(K^-p) &=& 11.93s^{0.0808}+25.33s^{-0.4525}
  \ \mathrm{mb}.
 \end{eqnarray}
  In both of the above, the first term corresponds to the pomeron
contribution. The second term is due to the $\rho/\omega/a/f$
trajectories. The difference in the two cases is due to the $C=-1$
trajectories, i.e., the $\rho$ and the $\omega$, contributing in the
opposite way. Thus the $a/f$ contribution that we are interested in can be
obtained as the mean of the two, i.e., 17.255. The same in the case of
$pp$ and $p\bar p$ cross sections gives 77.235.

  We then calculate the $\gamma K^\pm$ total cross section by assuming 
that the couplings to the trajectories factorize. We use our knowledge of 
the $\gamma K^\pm$, $\gamma p$, $pp$ and $p\bar p$ cross sections to 
write:
 \begin{eqnarray}
  \sigma_\mathrm{tot}(\gamma K^\pm) &\approx&
  \frac{11.93\times 0.0677}{21.7}s^{0.0808} +
  \frac{17.255\times 0.129}{77.235}s^{-0.4525} \ \mathrm{mb}\nonumber\\
  &=& 0.0372s^{0.0808}+0.0288s^{-0.4525} \ \mathrm{mb}.
 \end{eqnarray}
  This time, we obtain from eqn.~(\ref{eqn_ven_totcs_num}):
 \begin{eqnarray}
  \overline\beta(\gamma\gamma\to K^+K^-)\approx0.172.
 \end{eqnarray}

  There is an additional complication in the case of kaons because of the 
presence of the strange quark. We expect, as is evident in the data 
\cite{bellek}, a significant contribution from the $f'$ trajectory. The 
coupling of the $f'$ trajectory can only be estimated. For the sake of 
argument, let us say here that the coupling is proportional to the square 
of the electric charge of the strange quark, i.e., $e^2/9$. On the other 
hand, $f/a\sim(u\bar u+d\bar d)/\sqrt2$ so the average charge squared is 
2.5 times that of the strange quark. Hence we estimate $\overline\beta$ to 
be 2.5 times less than that of the $f/a$ trajectory. We shall see later 
that this leads to a not unreasonable description of the cross section.

  For the sake of completeness, let us calculate the case of $\pi^\pm$ 
production. In this case, taking the $\pi^\pm p$ scattering total cross 
sections as input, factorization yields:
 \begin{eqnarray}
  \sigma_\mathrm{tot}(\gamma \pi^\pm) &\approx&
  \frac{13.63\times 0.0677}{21.7}s^{0.0808} +
  \frac{31.79\times 0.129}{77.235}s^{-0.4525} \ \mathrm{mb}\nonumber\\
  &=& 0.0425s^{0.0808}+0.0531s^{-0.4525} \ \mathrm{mb}.
 \end{eqnarray}
  Hence, again from eqn.~(\ref{eqn_ven_totcs_num}) and assuming that the 
slope of the pion trajectory is also 0.9, we obtain:
 \begin{eqnarray}
  \overline\beta(\gamma\gamma\to \pi^+\pi^-)=0.318.
 \end{eqnarray}

 \subsection{The additive quark rule} \label{sec_additive_quark}

  When considering hadrons other than $p,n,\pi^\pm$ and $K^\pm$, the above 
prescription fails due to the lack of data. In this case, the best that we 
can do is to estimate the coefficients by the additive quark rule.

  As noted in ref.~\cite{dlcs}, the additive quark rule is a good 
rule-of-the-thumb for the pomeron. We can say that the pomeron couples 
with equal strengths to the $u$-quark and the $d$-quark, and with 70\% of 
the strength to the $s$-quark. This gives good estimation for $p\pi^\pm$ 
and $p K^\pm$ scattering cross sections. However, this does not work for 
the $a/f$ couplings.

  For example, we see from the above that $\overline\beta$ is about 0.318
for $\gamma\gamma\to\pi^+\pi^-$ and is about 0.172 for $\gamma\gamma\to
K^+K^-$. Since there are two $u/d$ quarks in the pion and only one in the
kaon, it seems that $0.318/0.172\sim2$ is in accordance with expectation.
The problem is that from the same line of thought, we would expect
$\overline\beta$ for $\gamma\gamma\to p\bar p$ to be about 1.5 times the
pionic case, and this is incorrect as we have 0.770 in this case rather 
than $\sim0.48$.

  However, for the sake of rough estimation, we can proceed by applying 
the additive quark rule separately to the mesons and the baryons. We then
have, for baryons:
 \begin{equation}
  \overline\beta(\gamma\gamma\to\Lambda\overline\Lambda) \approx
  \overline\beta(\gamma\gamma\to\Sigma\overline\Sigma) \approx
  \frac23 \overline\beta(\gamma\gamma\to p\bar p).
  \label{eqn_additive_quark_estimate}
 \end{equation}
  In our approximation, for the case of $\Sigma$, only the $f$ trajectory,
and not the $a$ trajectory, contributes.

  We should additionally consider the coupling of the $f'$ trajectory to 
the $s$-quark content of $\Lambda$ and $\Sigma$. We can again make a rough 
estimate of this by considering the squared electric charge of the 
$s$-quark, and say that $\overline\beta$ in this case is about one fifth 
of the $\overline\beta$ for the $f$ trajectory. We assume that the 
trajectory of eqn.~(\ref{eqn_traj_fprime}) holds also at the baryonic mass 
scale.

  One possibly severe drawback of this approach is that we have not taken 
into account the difference in isospin between $\Lambda$, with $I=0$, the 
proton, with $I=1/2$, and $\Sigma^0$, with $I=1$. In particular, $\Lambda$ 
is exchange anti-symmetric between $u$ and $d$ so we expect that the 
exchange symmetric, i.e., isospin-0, $f$ trajectory should not couple to 
the $\Lambda$. We know that the isospin-1 $a$ trajectory has much smaller 
coupling \cite{ddln} to $p$ and $n$ than the $f$ trajectory, so that we 
would expect the $\gamma\gamma\to\Lambda\overline\Lambda$ cross section to 
be much smaller than is implied by 
eqn.~(\ref{eqn_additive_quark_estimate}).

  We must turn to the data in order to resolve this point. The data from 
L3 \cite{baryon_l3} seem to suggest that the $\Lambda$ pair production 
cross section, at the level of the total cross section, is comparable to 
the $\Sigma^0$ pair production cross section. However, we should wait for 
improved statistics in order to confirm this point, and we should note 
furthermore that there is possibly discrepancy in the data on $\Lambda$ 
pair photoproduction between L3 \cite{baryon_l3} and CLEO 
\cite{baryon_cleo} which needs to be understood.

  Let us here proceed by taking the additive quark rule literally as the 
rule-of-the-thumb governing the sum of the couplings of the $f$ and the 
$a$ trajectories. As we shall see in the following section, this in fact 
leads to a greater total cross section for $\Lambda$ than $\Sigma$, 
indicating the need to take into account the suppression due to isospin.

  There is no difference between $\Sigma^-,\Sigma^0$ and $\Sigma^+$. Since
their masses and the trajectories are also similar, we have:
 \begin{equation}
  \sigma(\gamma\gamma\to\Sigma^+\overline{\Sigma^+})\approx
  \sigma(\gamma\gamma\to\Sigma^0\overline{\Sigma^0})\approx
  \sigma(\gamma\gamma\to\Sigma^-\overline{\Sigma^-}).
 \end{equation}
  As mentioned above, the same relation holds between the proton and the 
neutron cross sections.
  These relations are independent of the additive quark rule, and are also
independent of the formalism that we have adopted.
  Any violation of these equalities would indicate that there is 
significant contribution from the isospin-1 resonances, and this would 
seem implausible from the similarity of the $pn,\bar pn$ total cross 
sections with the $pp,\bar pp$ total cross sections.

  We note that another possible source of violation of this flavour
symmetry would be the third term of fig.~\ref{fig_duality}, which is also
expected to be small except in the forward region.

  Our predictions are thus in contrast with those of 
ref.~\cite{berger-schweiger}, in which the exchange symmetry is between 
the $d$- and the $s$-quarks rather than the $u$- and the $d$-quarks.

 \section{Result and discussions}\label{sec_result}

  Following the above discussions of the formalism and parametrization, 
let us now proceed to examining the results of our calculation for the 
cases of $\gamma\gamma\to p\bar p$ and $\gamma\gamma\to K^+K^-$, after 
which we turn to the case of the other baryons.

 \subsection{$\gamma\gamma\to p\bar p$}

 \FIGURE[ht]{
 \epsfig{file=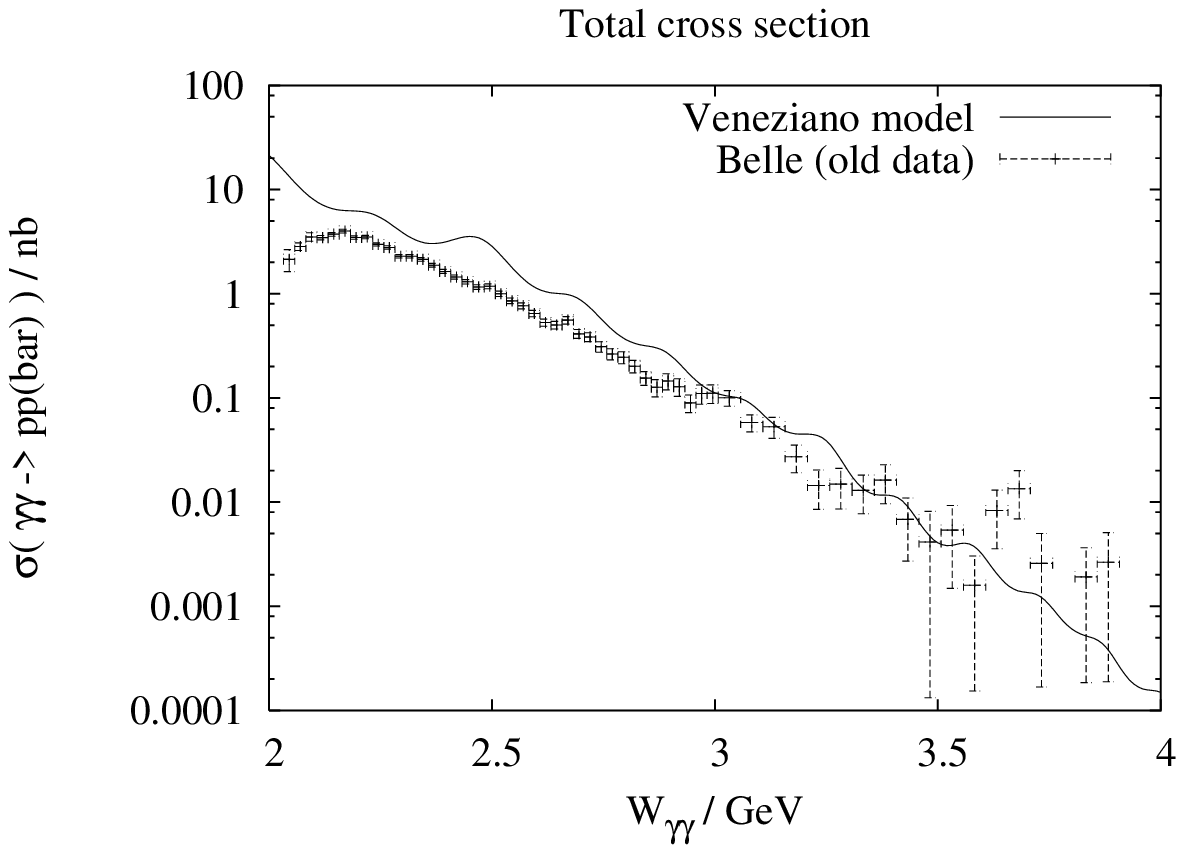,width=14cm}
 \caption{The $\gamma\gamma\to p\bar p$ total cross section in the region
$|\cos\theta^*|<0.6$, calculated using the Veneziano model.
  The Belle result is shown for comparison. The vertical error-bars 
represents the statistical uncertainty.
 \label{fig_ven_cs}}}

 \FIGURE[ht]{
 \epsfig{file=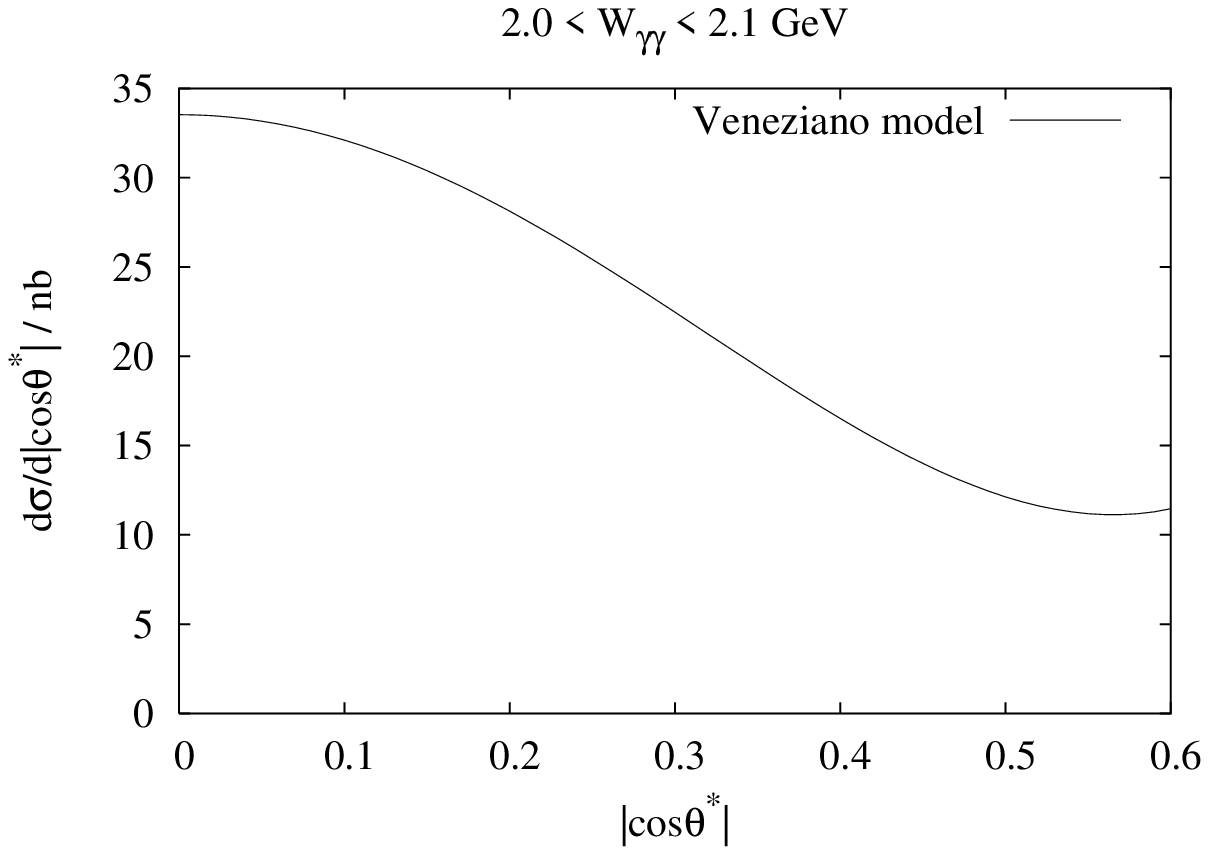,width=4.9cm}
 \epsfig{file=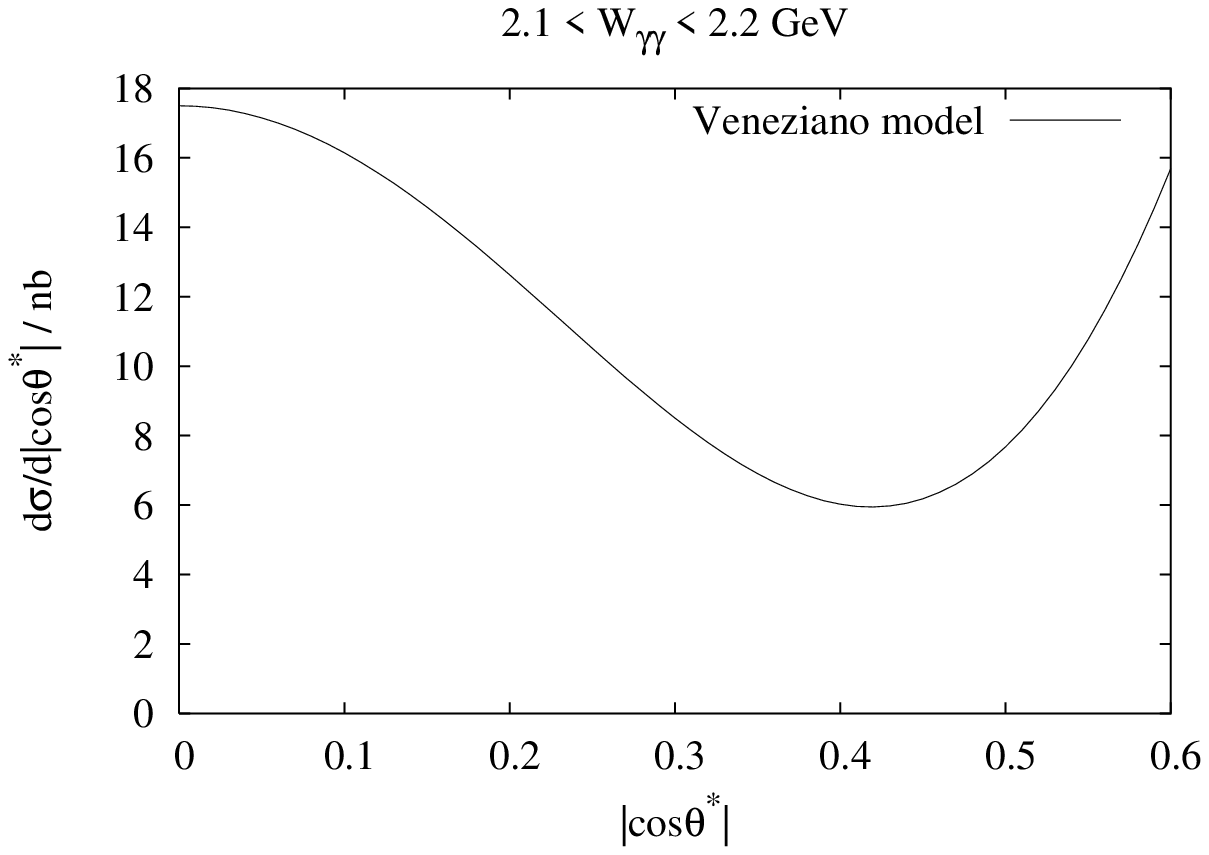,width=4.9cm}
 \epsfig{file=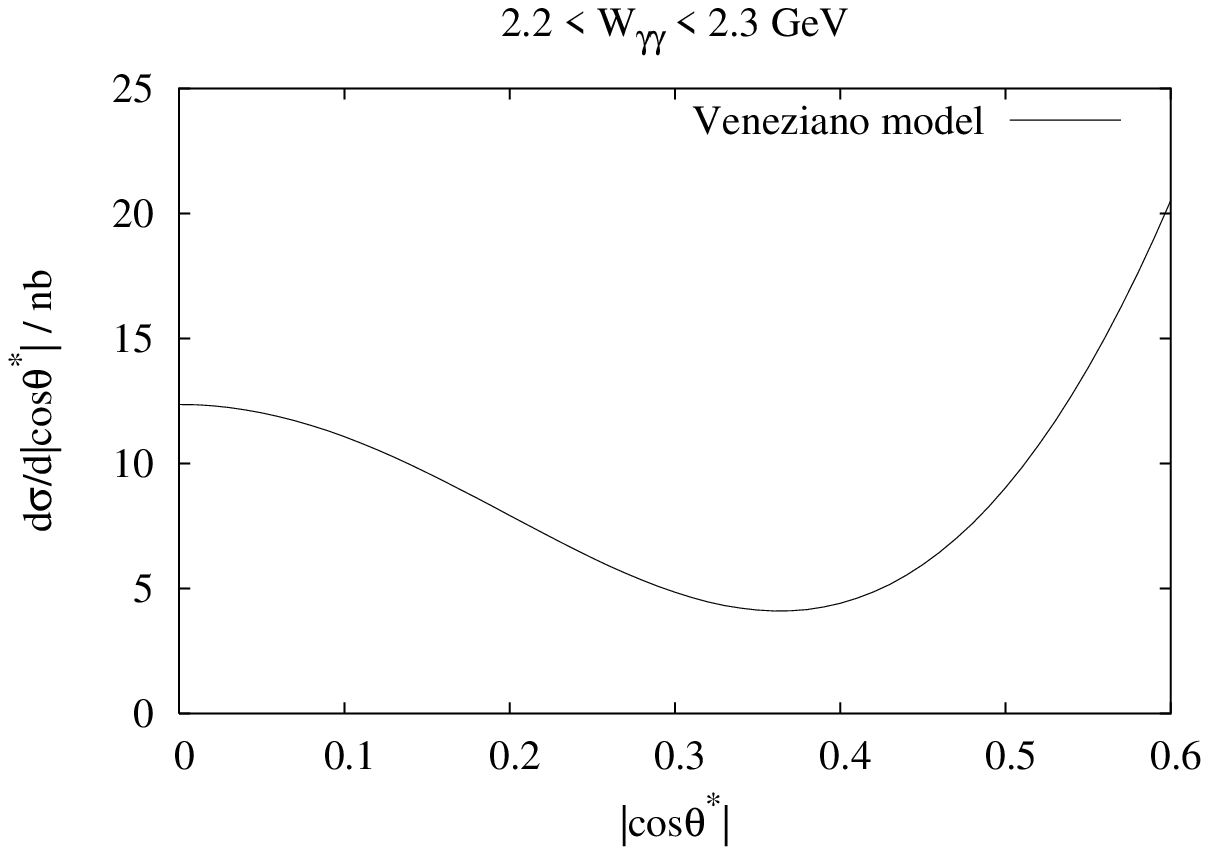,width=4.9cm}

 \epsfig{file=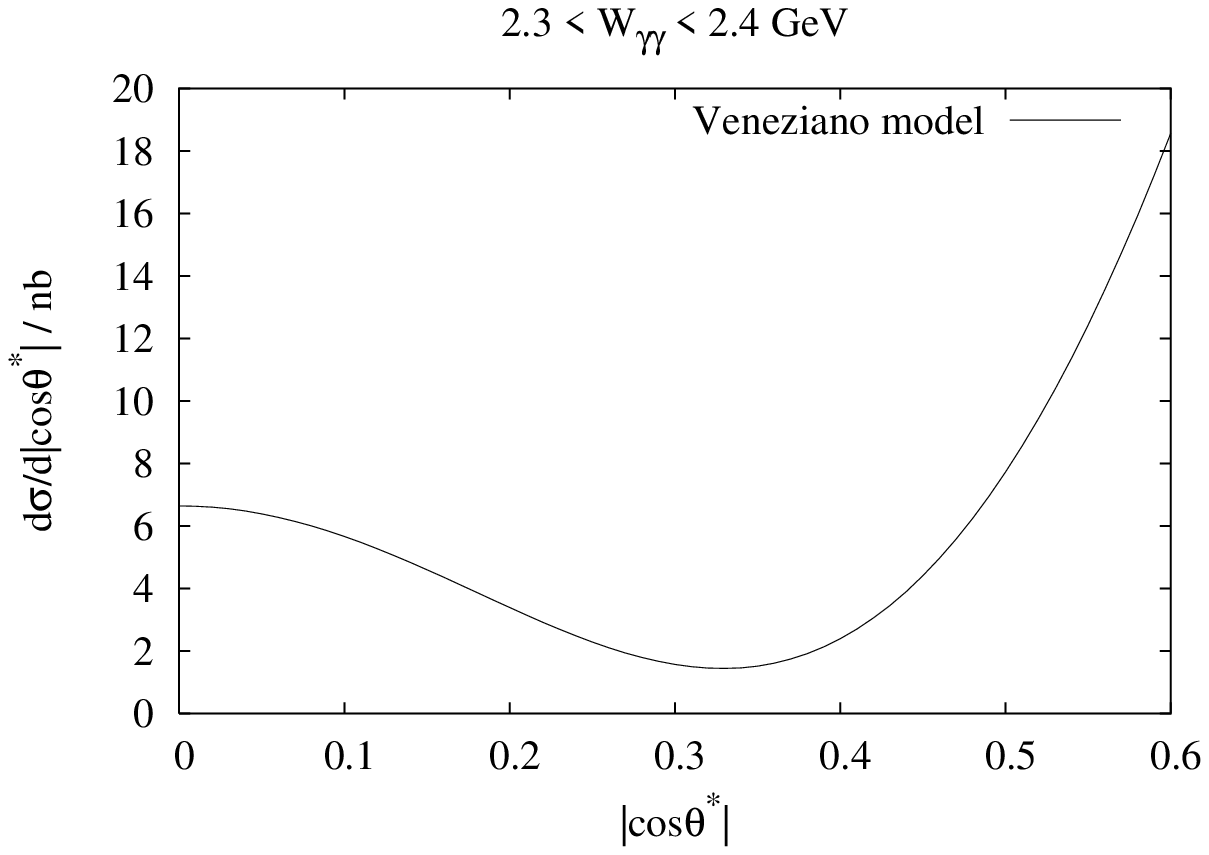,width=4.9cm}
 \epsfig{file=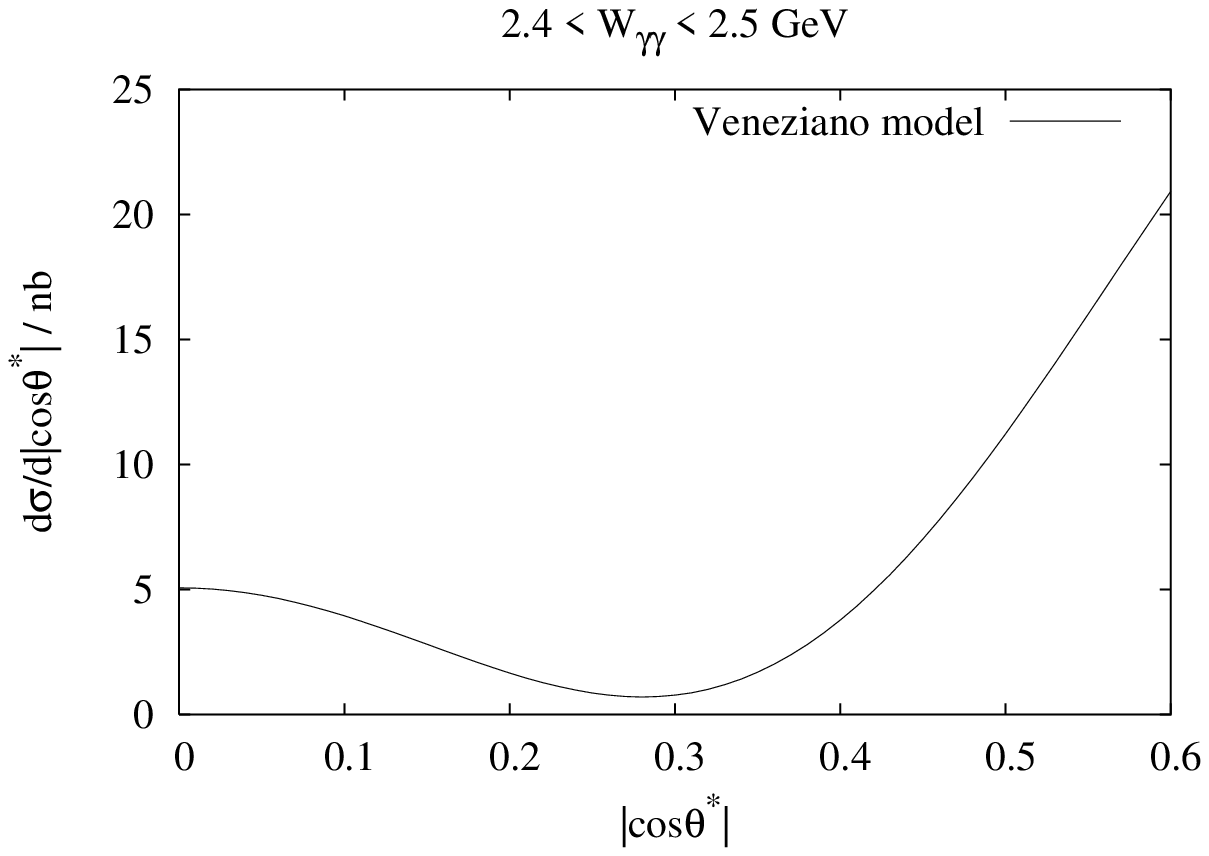,width=4.9cm}
 \epsfig{file=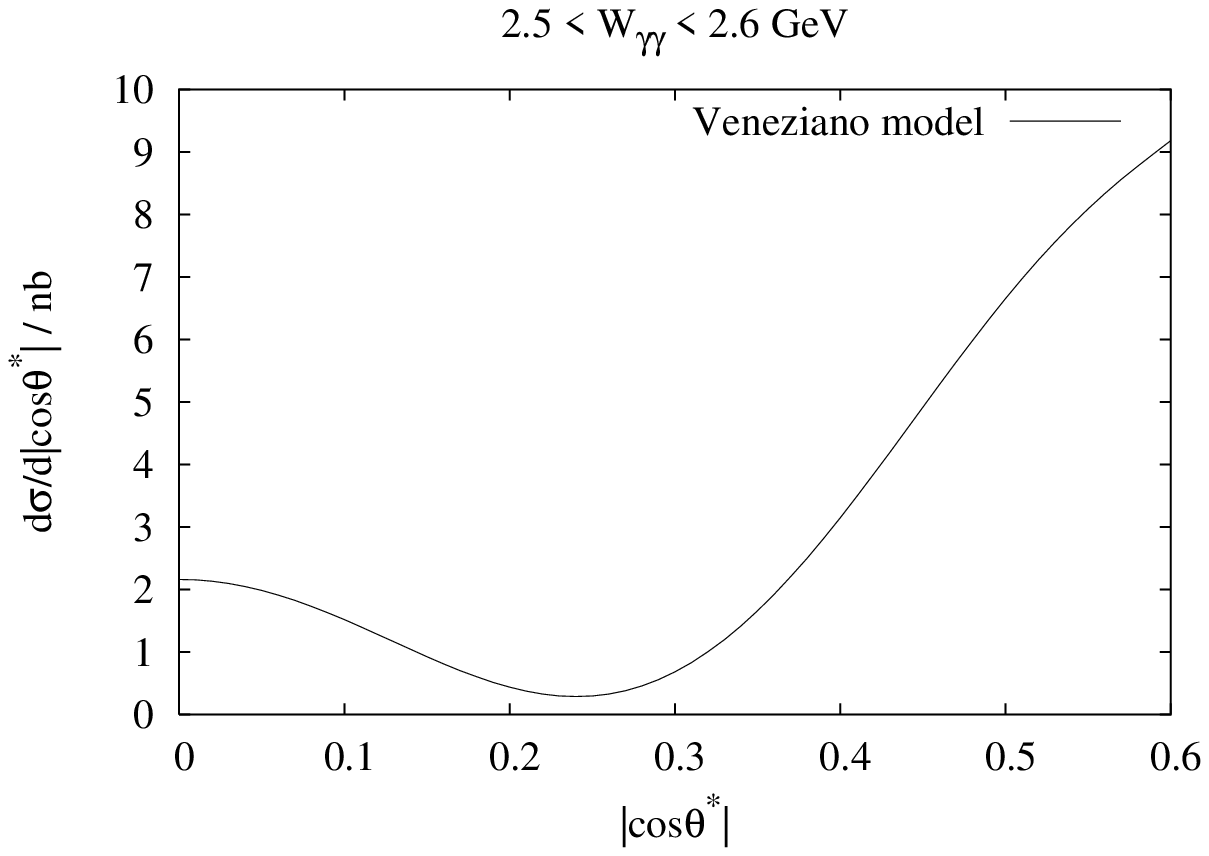,width=4.9cm}

 \epsfig{file=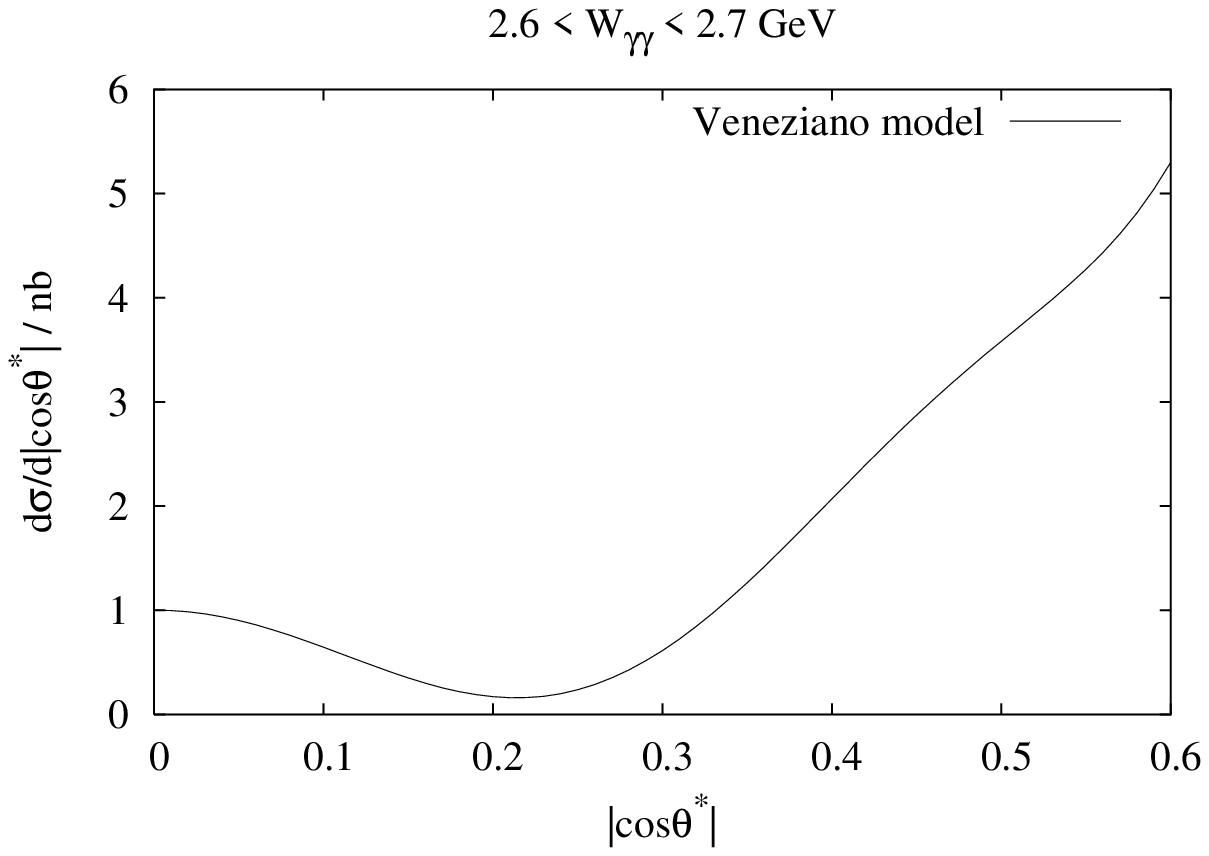,width=4.9cm}
 \epsfig{file=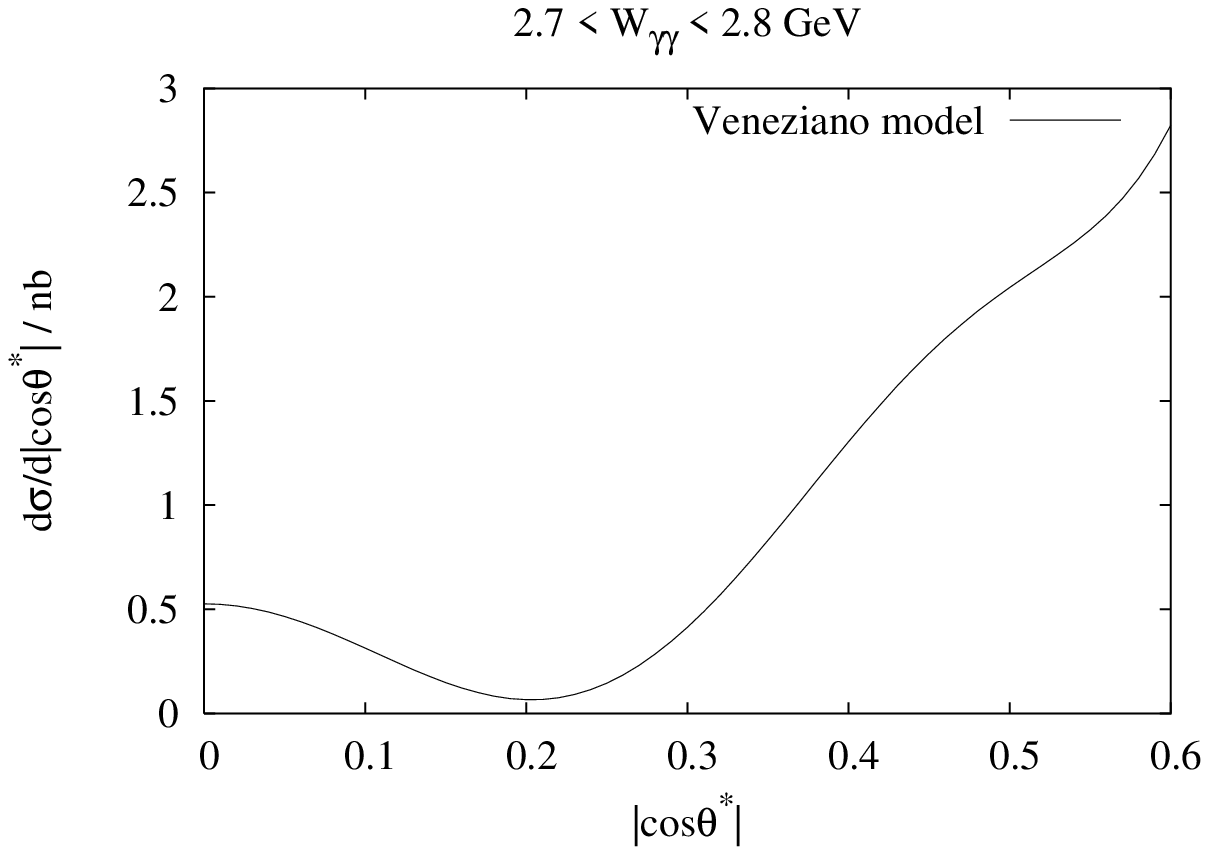,width=4.9cm}
 \epsfig{file=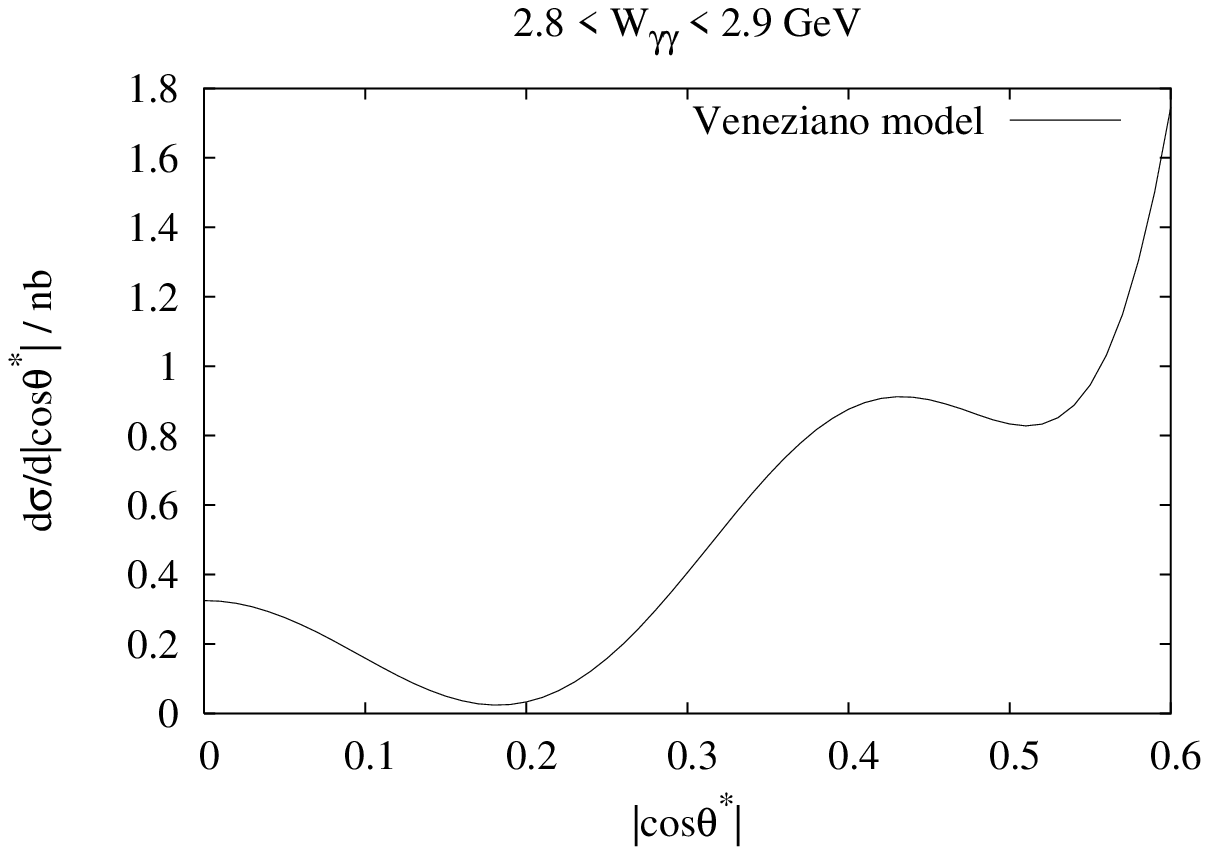,width=4.9cm}

 \epsfig{file=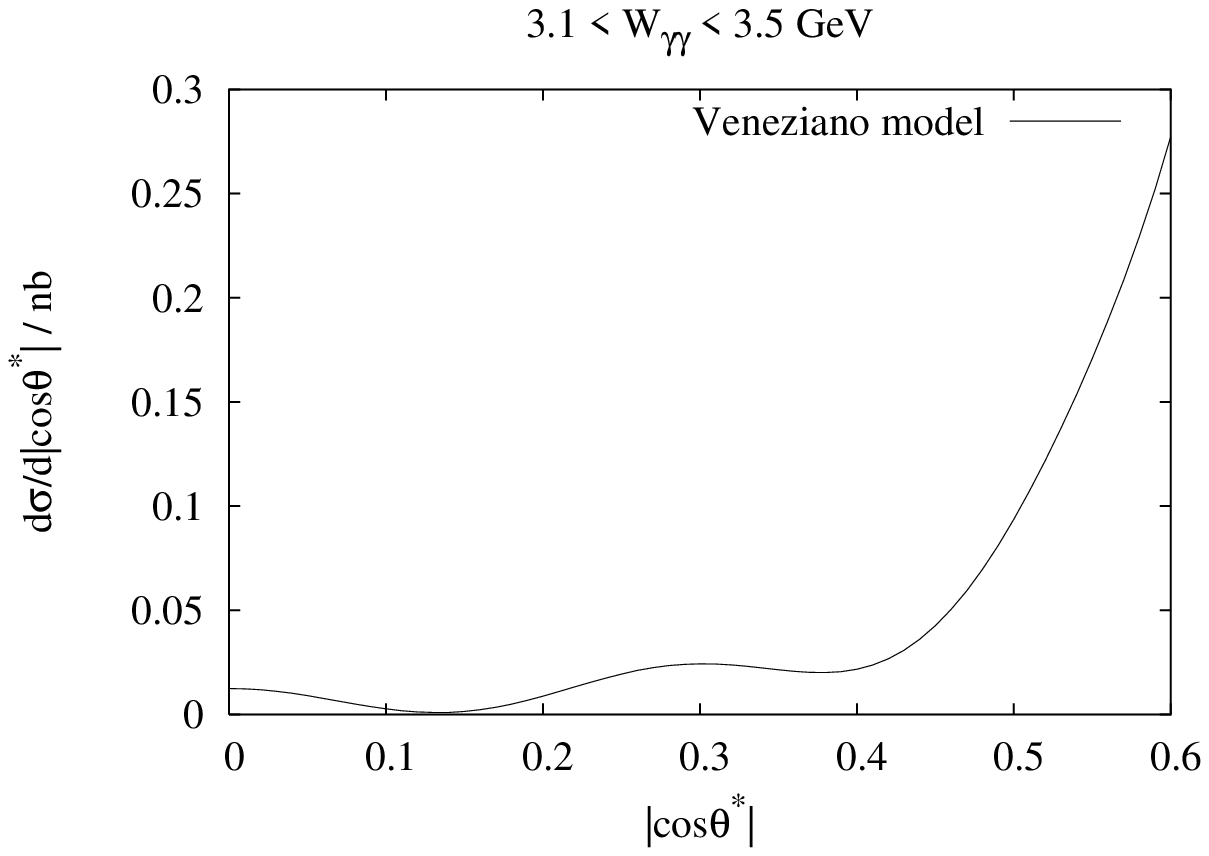,width=4.9cm}
 \epsfig{file=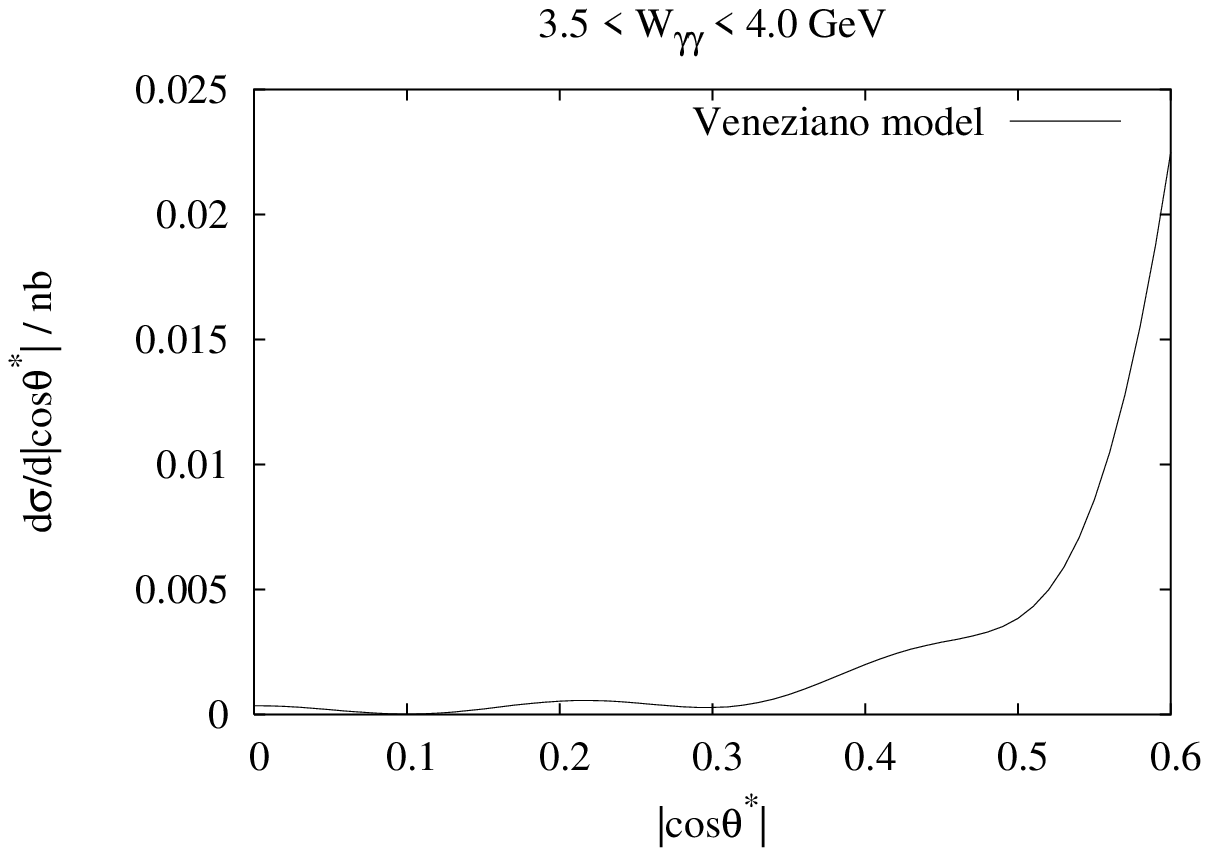,width=4.9cm}
 \caption{The angular distribution for $\gamma\gamma\to p\bar p$.
 \label{fig_ven_ad}}}

  We first show the result of calculation for the $\gamma\gamma\to p\bar 
p$ cross section in figs.~\ref{fig_ven_cs} and \ref{fig_ven_ad}.
  The Belle result shown in fig.~\ref{fig_ven_cs} corresponds to the old 
data points of ref.~\cite{photon03}. The more recent data points from 
ref.~\cite{bellep} exhibit slightly higher cross sections, and so are more 
in agreement with the calculation.

  The total cross section, shown in fig.~\ref{fig_ven_cs}, shows a good
$W_{\gamma\gamma}$ dependence up to about 3 GeV. This is as expected as
perturbative contributions are expected to become increasingly dominant at
higher energies and 3 GeV seems to be a reasonable turnover point.

  We would expect that the turnover point is determined not only by 
$W_{\gamma\gamma}$ but also by $\cos\theta^*$ because the relevant hard 
scale 
for QCD should be approximately the transverse momentum $p_T$. In the more 
central region, we would expect that the perturbative contributions set in 
earlier, and the agreement of our result with the measurement would be 
poorer.

  The resonances are more pronounced in this analysis than the real data.  
However, the positions of the resonances seem to match some of the 
fluctuations in the real data. There are extra contributions in the real 
data, for instance, from a peak corresponding to $\eta_c(2980)$ which is, 
of course, absent in the present analysis.

  The threshold behaviour near 2 GeV seems wrong in the present
calculation. We note that this is not definitive as the experimental
statistics in this region is small, but we have a reasonable understanding
as to why this is the case, which is as follows.

  We know that the interactions of both the gluon and the photon with 
quarks are chirality conserving. Hence in the limit where the masses are 
small compared to the centre-of-mass energy, the interactions would 
predominantly conserve helicity. If this holds to some degree even in the 
threshold region, then the final state proton helicity would be opposite 
to the antiproton helicity here also. The total $s$-channel helicity along 
the direction of the proton would thus be $\pm1$. This can not be mediated 
by a spin-0 particle. The problem is that the amplitude as written in 
eqn.~(\ref{eqn_baryonamp}) contains contributions from spin-0 daughter 
resonances.

  Just above the threshold, spin-0 terms in eqn.~(\ref{eqn_baryonamp}),
i.e., the $s$-wave terms, are enhanced compared to the higher-spin terms
by virtue of having no suppression by integer powers of $(1-4m_p^2/s)$
where $m_p$ is the proton mass. This thus explains why the threshold
region is described particularly badly in this approach. This also
provides one possibility as to why the calculated cross section is
higher than the real data below $W_{\gamma\gamma}\approx 3$ GeV.

  This hypothesis can be investigated simply by analyzing the angular 
distribution in the threshold region\footnote{We should note that spin-0 
terms are not completely excluded as, for instance, the $\eta_c(2980)$ 
resonance is spin-0. We thank C.C.~Kuo for reminding us of this point.}.

  Turning now to the angular distributions shown in fig.~\ref{fig_ven_ad},
we see that a good qualitative description of the general behaviour is
obtained.
  Unfortunately the positions of the dips are not in exact agreement with
data.

  As a general trend, the $t$-channel behaviour sets in earlier than in 
the data. This is presumably due to the combination of the two 
deficiencies of the model mentioned earlier, namely that the baryonic 
Veneziano amplitude of eqn.~(\ref{eqn_baryonamp}) has wrong forward 
behaviour in the Regge region, and that baryon exchange processes or 
photoproduction processes are in any case not described well by a simple 
Regge pole picture.

  Nevertheless, the numbers seem more than satisfactory for a simple model 
with known deficiencies and no adjustable parameters.

  The secondary dip seen above 2.8 GeV in the model is an interesting 
feature that merits further study with increased experimental statistics. 
Physically, this is due to the presence of resonances with higher $J$.

  The dips seem more exaggerated in the Veneziano model compared with the 
real data. One explanation is the greater smearing between the peaks found 
in the real data in the plot of the total cross section, which in turn 
mixes the contributions from different resonances.

  Another possible explanation is that there is further contribution from 
helicity-flipped contributions. As noted just below 
eqn.~(\ref{eqn_optical}), we only consider the cases in which the photon 
and the hadron helicities are separately conserved. This is likely to be a 
good approximation for the proton, but for the photon, there is presumably 
further contribution from the $(\pm,\mp)$ helicity combinations. These do 
not mix with the $(\pm,\mp)$ helicity combinations considered here, and so 
the resulting amplitude would add in quadrature with our amplitude. Hence 
the dips would be less prominent.

  This point would be interesting to test experimentally, though unlikely 
in the existing experimental environment. A more plausible quantity to 
measure would be the helicity of the baryons in the pair production of 
unstable baryons. This would be interesting and the measurement should 
shed light onto some of the points discussed above, for which we do not 
yet have a conclusive statement.

 \subsection{$\gamma\gamma\to K^+K^-$}

  Let us now turn our attention to the $\gamma\gamma\to K^+K^-$ process.

 \FIGURE[ht]{
 \epsfig{file=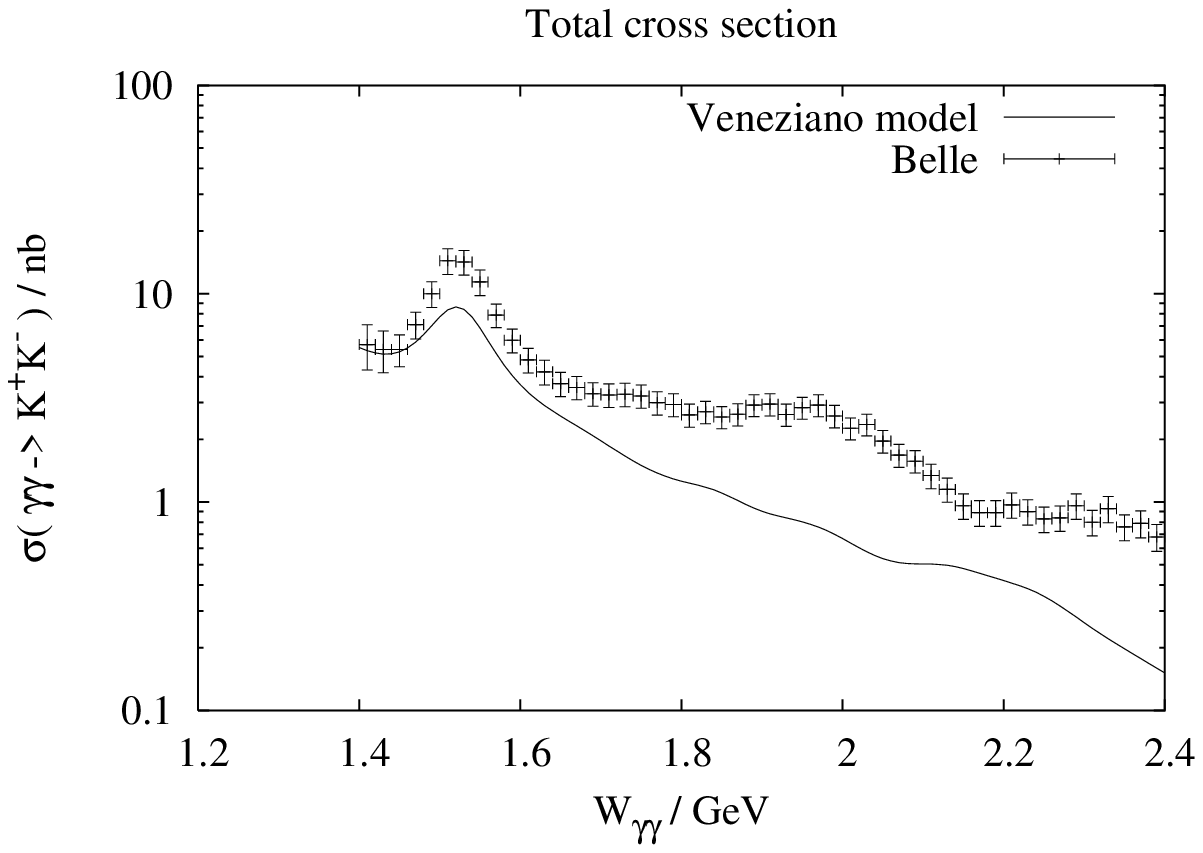,width=14cm}
 \caption{The $\gamma\gamma\to K^+K^-$ total cross section, in the region 
$|\cos\theta^*|<0.6$, calculated using the Veneziano model. The vertical 
error-bars on the Belle data-points represent the sum in quadrature of the 
statistical and systematic errors.
 \label{fig_kpkm_ven_cs}}}

 \FIGURE[ht]{
 \epsfig{file=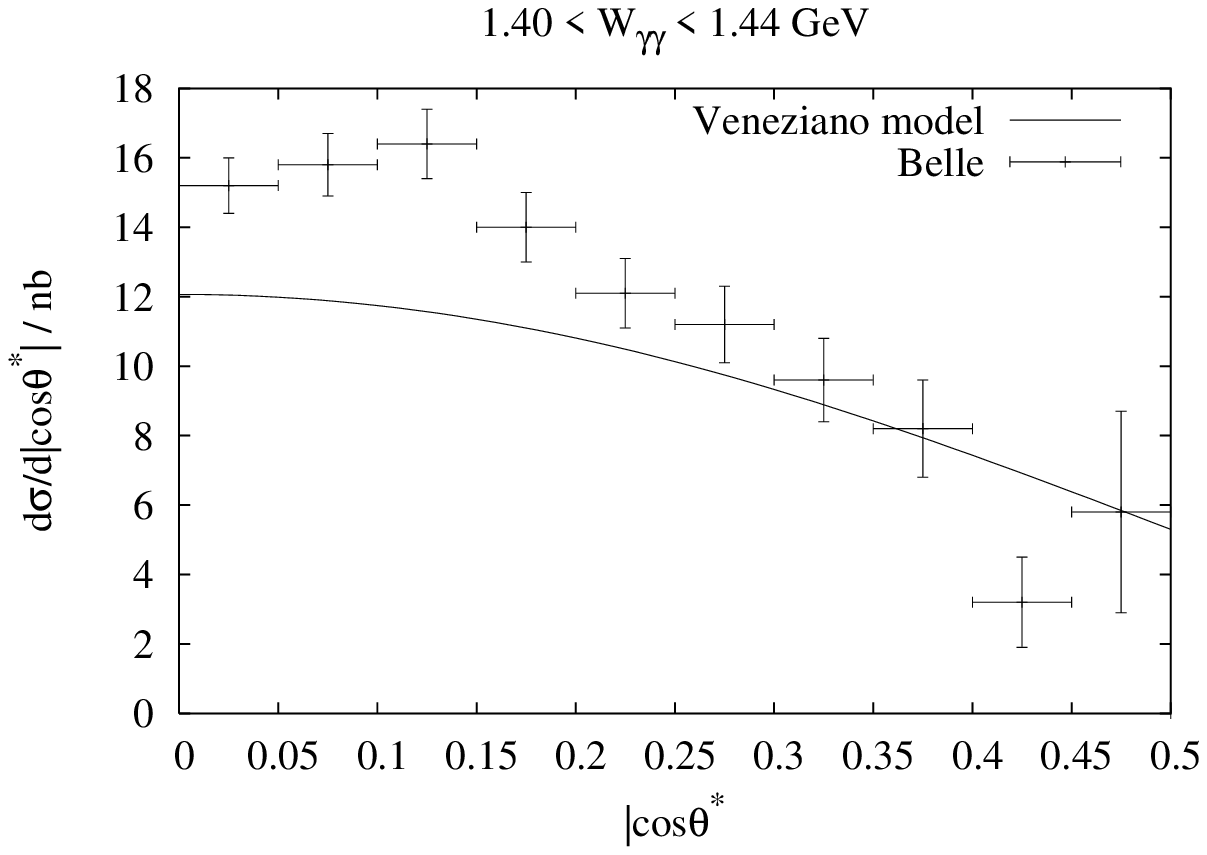,width=3.3cm}
 \epsfig{file=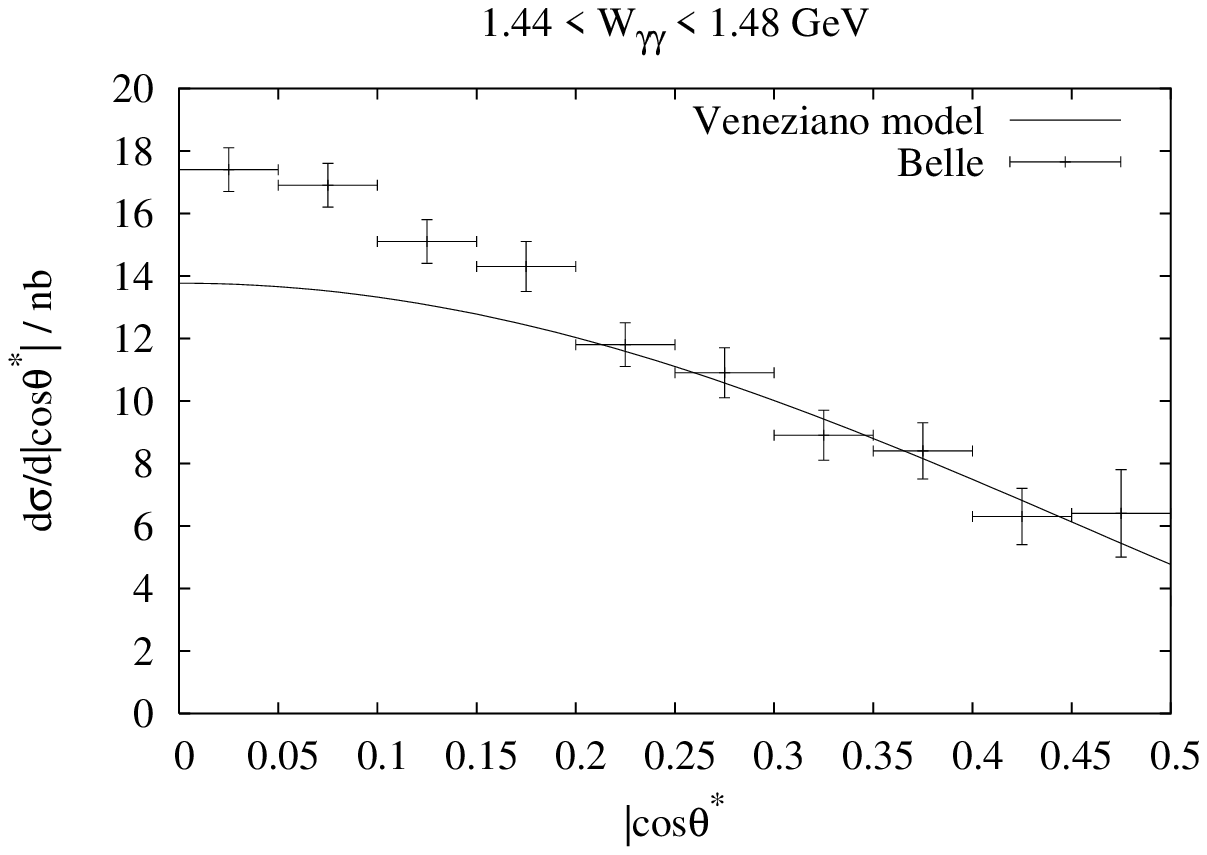,width=3.3cm}
 \epsfig{file=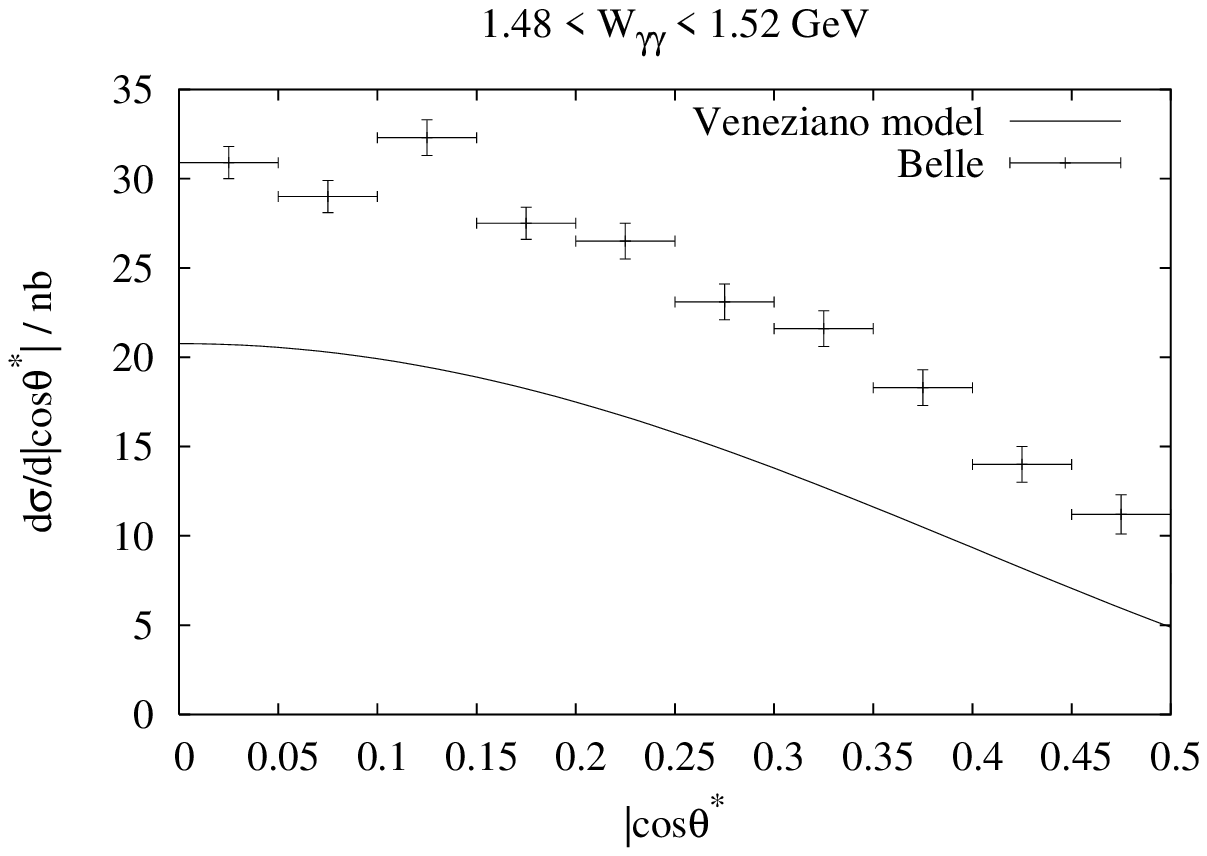,width=3.3cm}
 \epsfig{file=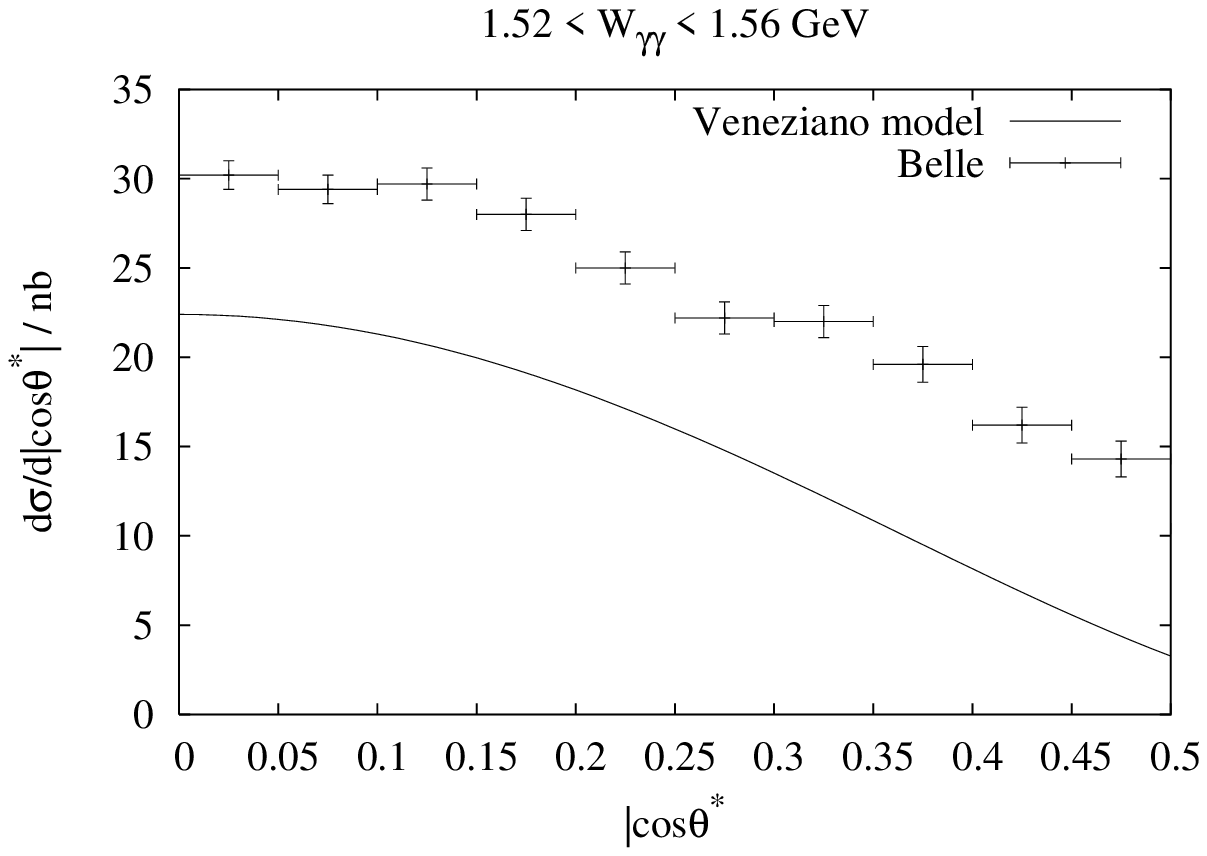,width=3.3cm}

 \epsfig{file=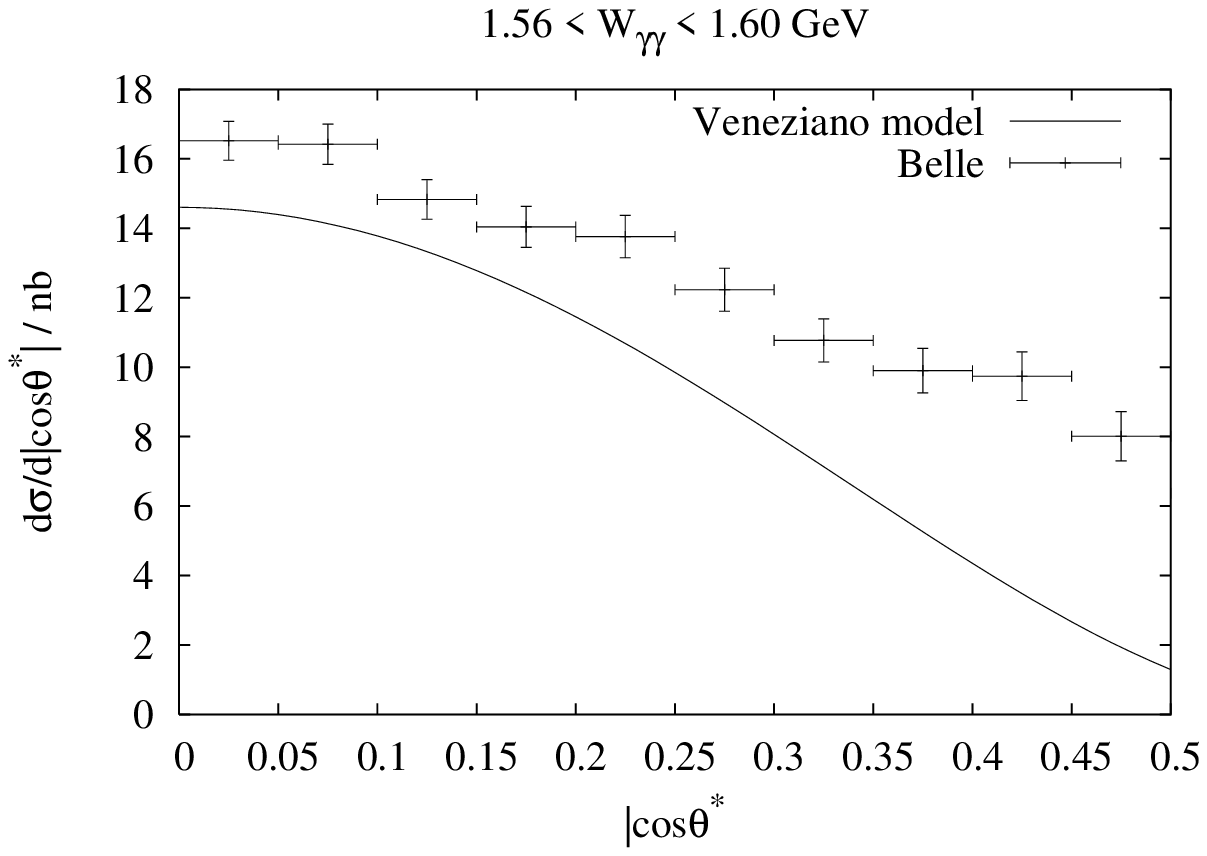,width=3.3cm}
 \epsfig{file=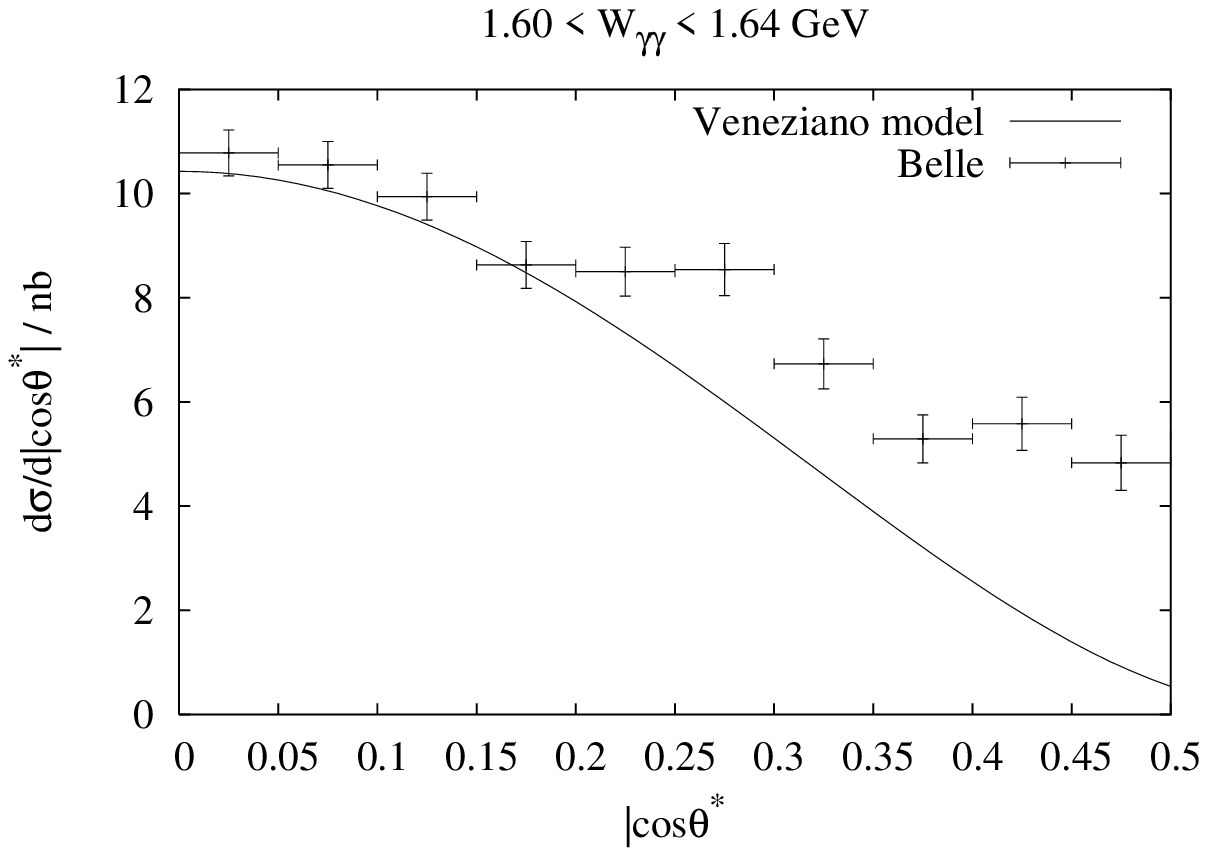,width=3.3cm}
 \epsfig{file=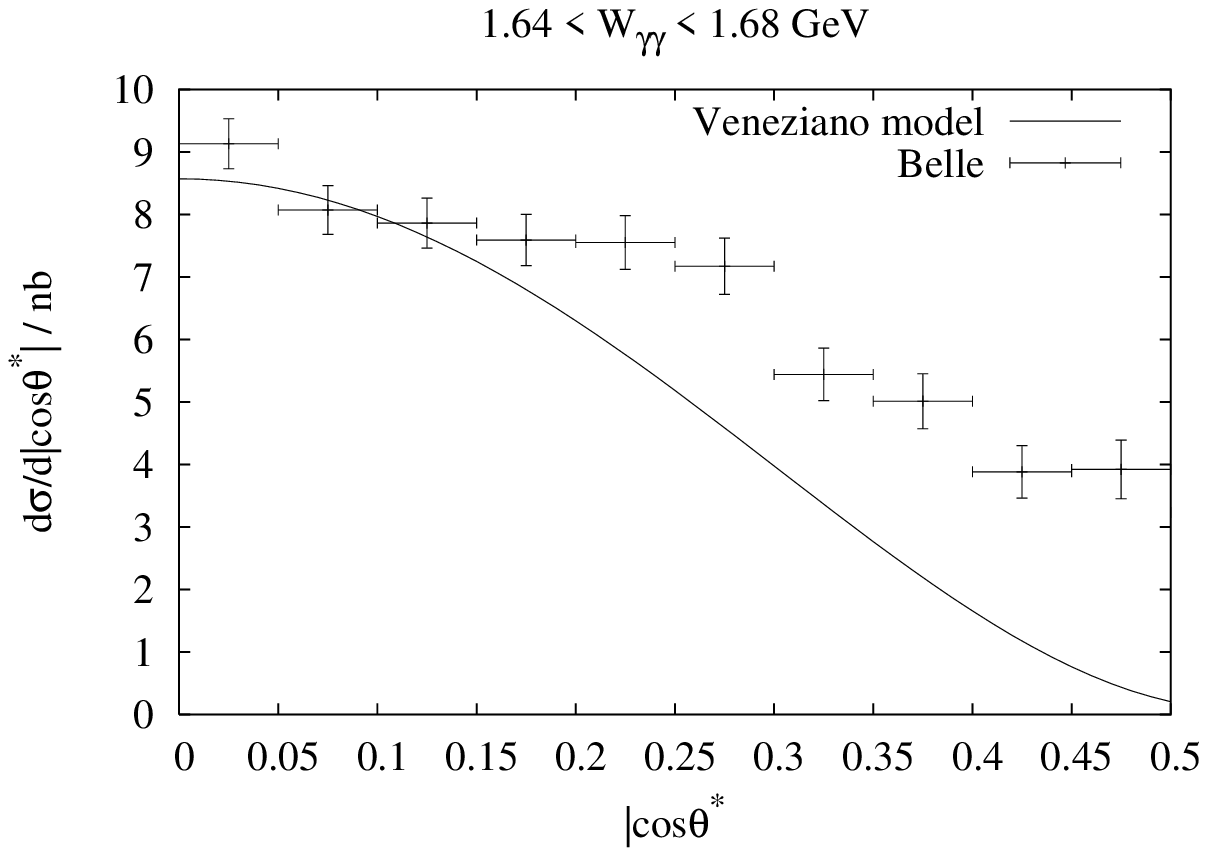,width=3.3cm}
 \epsfig{file=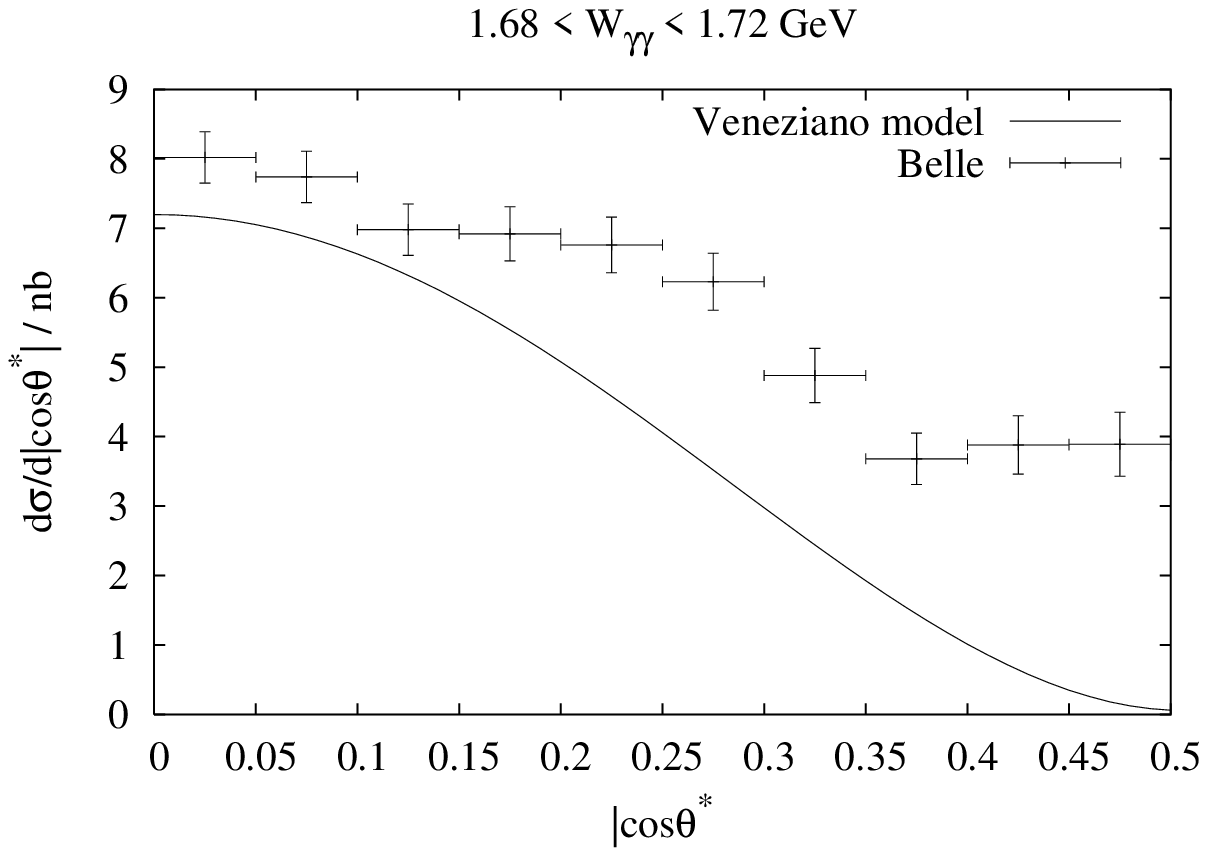,width=3.3cm}

 \epsfig{file=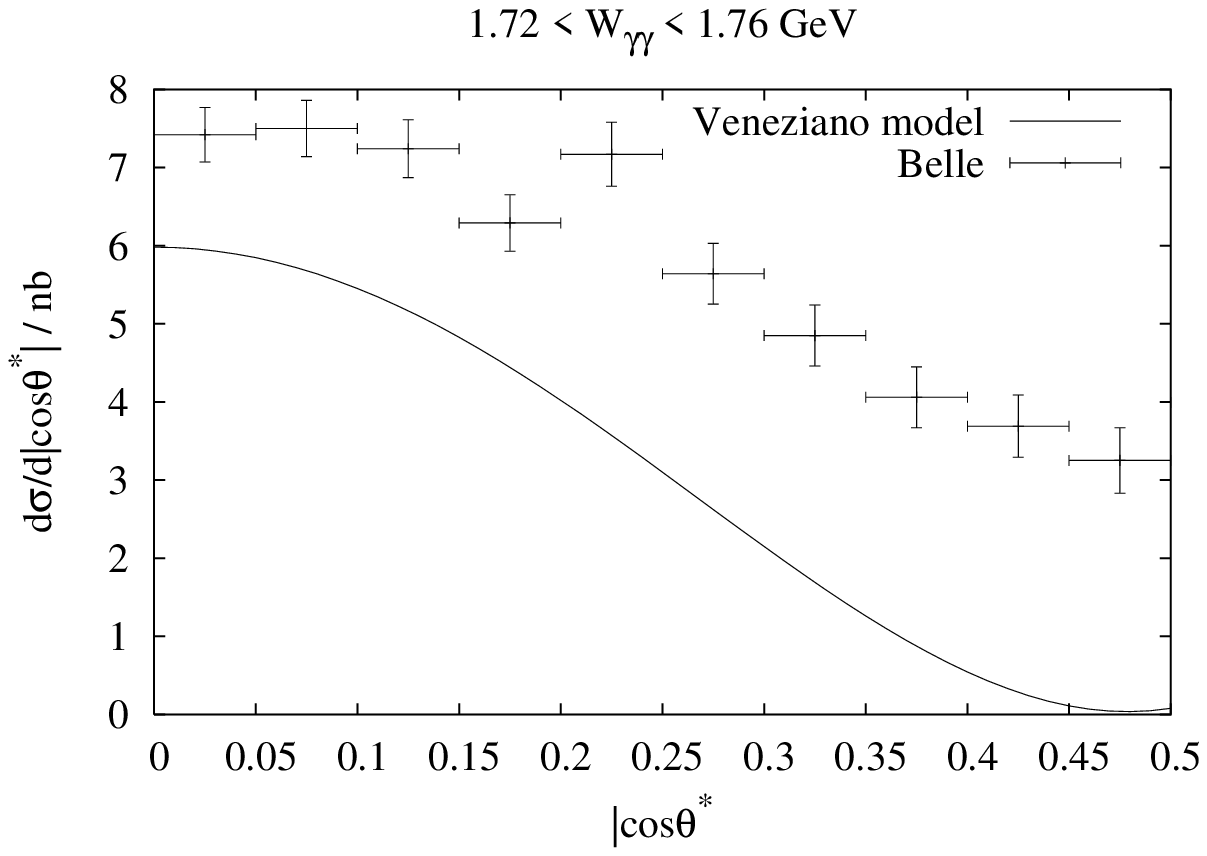,width=3.3cm}
 \epsfig{file=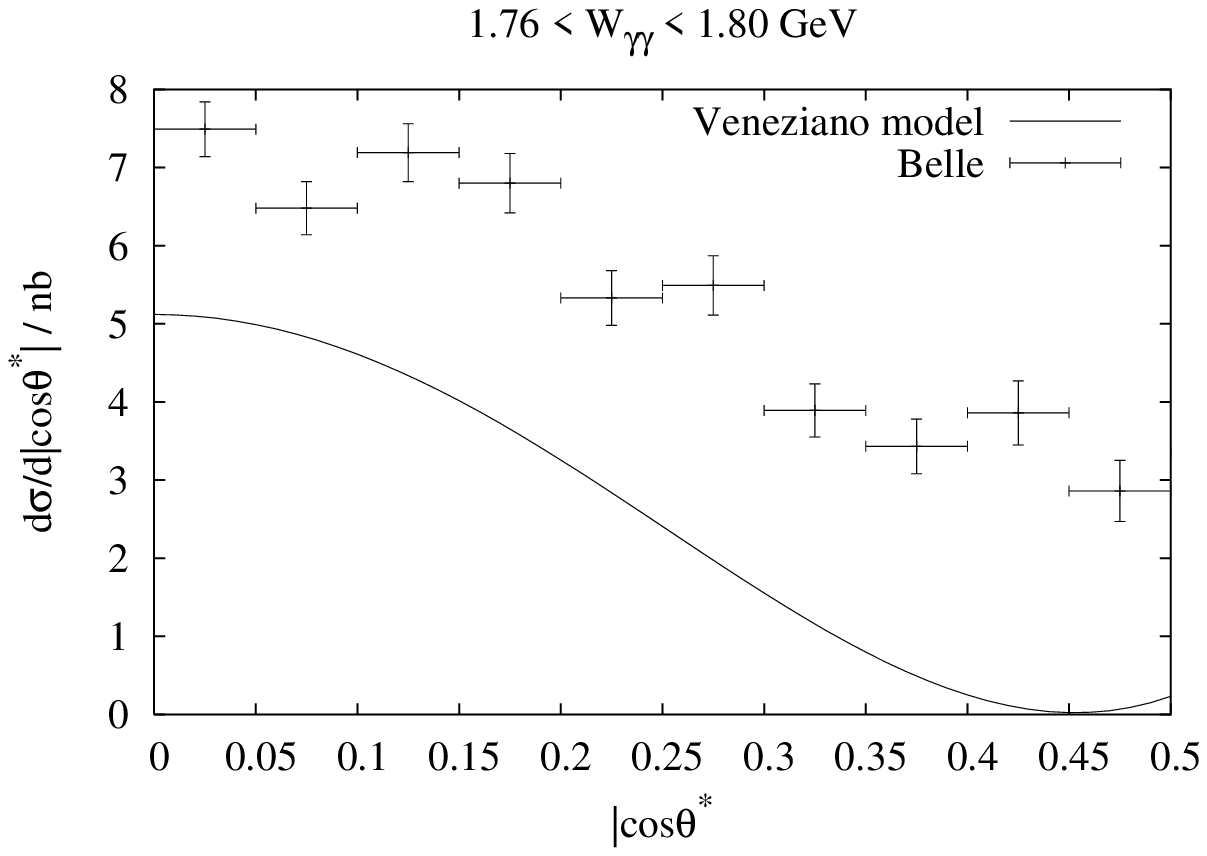,width=3.3cm}
 \epsfig{file=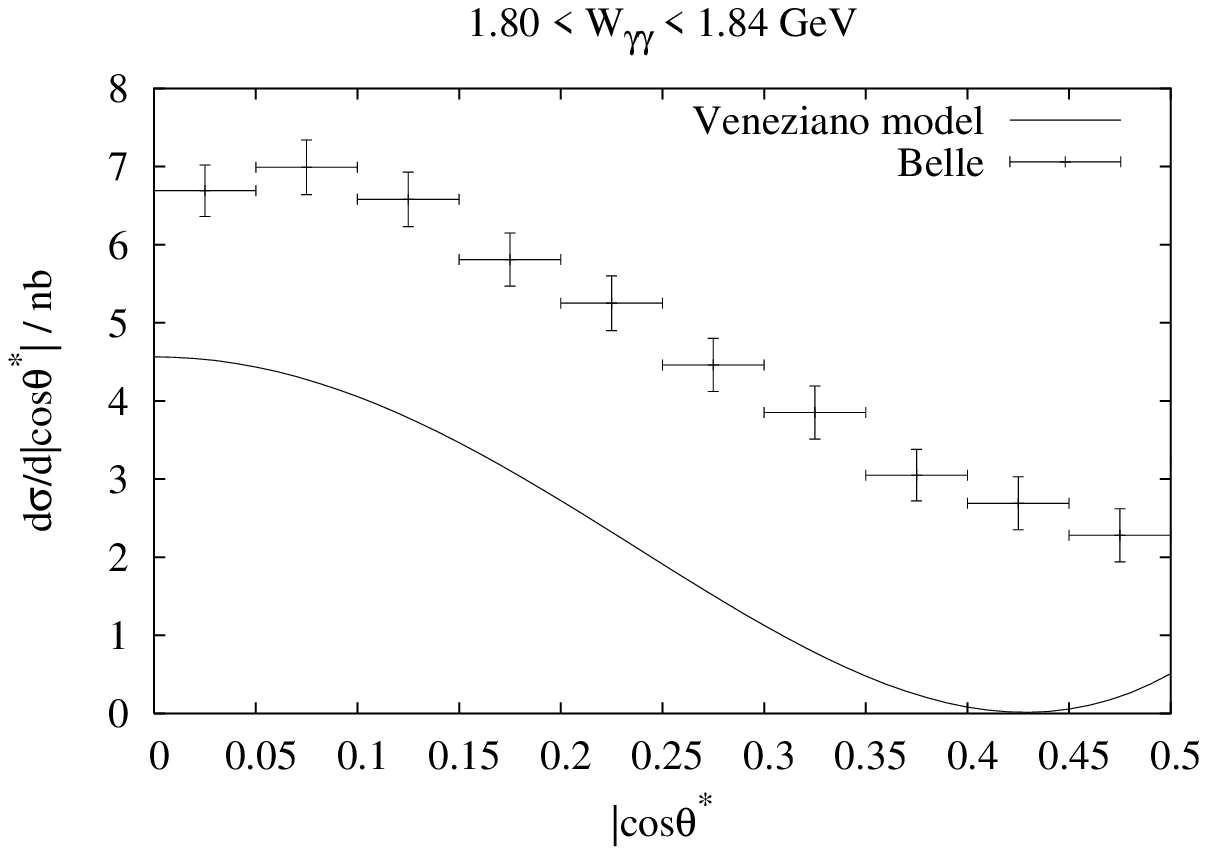,width=3.3cm}
 \epsfig{file=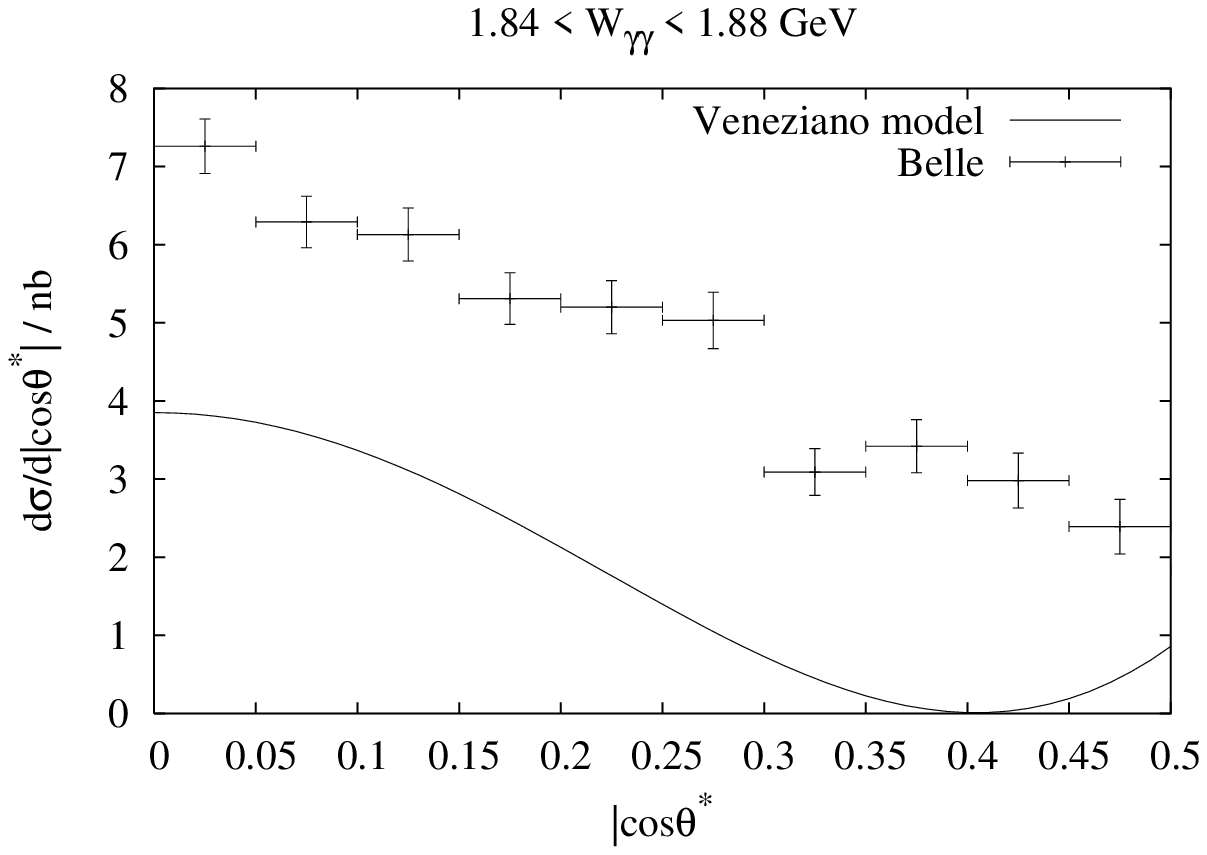,width=3.3cm}

 \epsfig{file=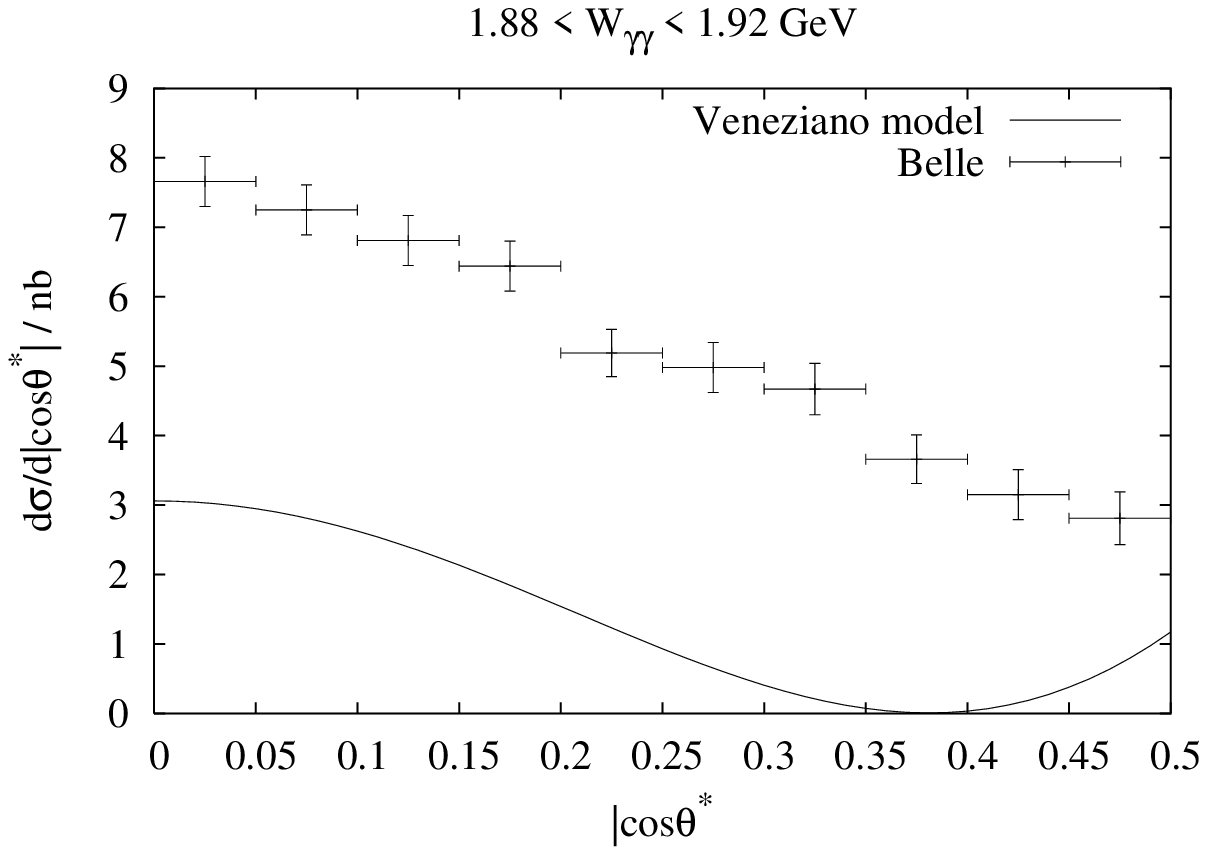,width=3.3cm}
 \epsfig{file=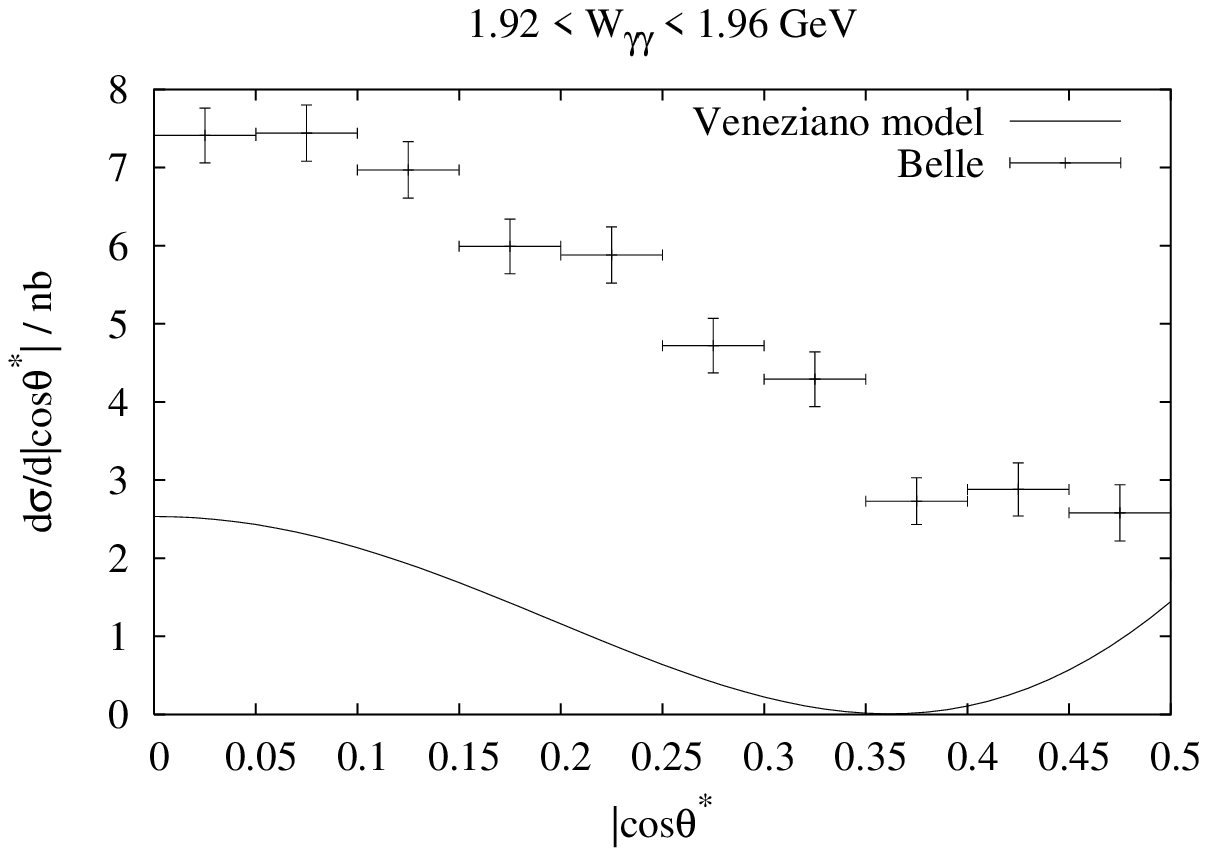,width=3.3cm}
 \epsfig{file=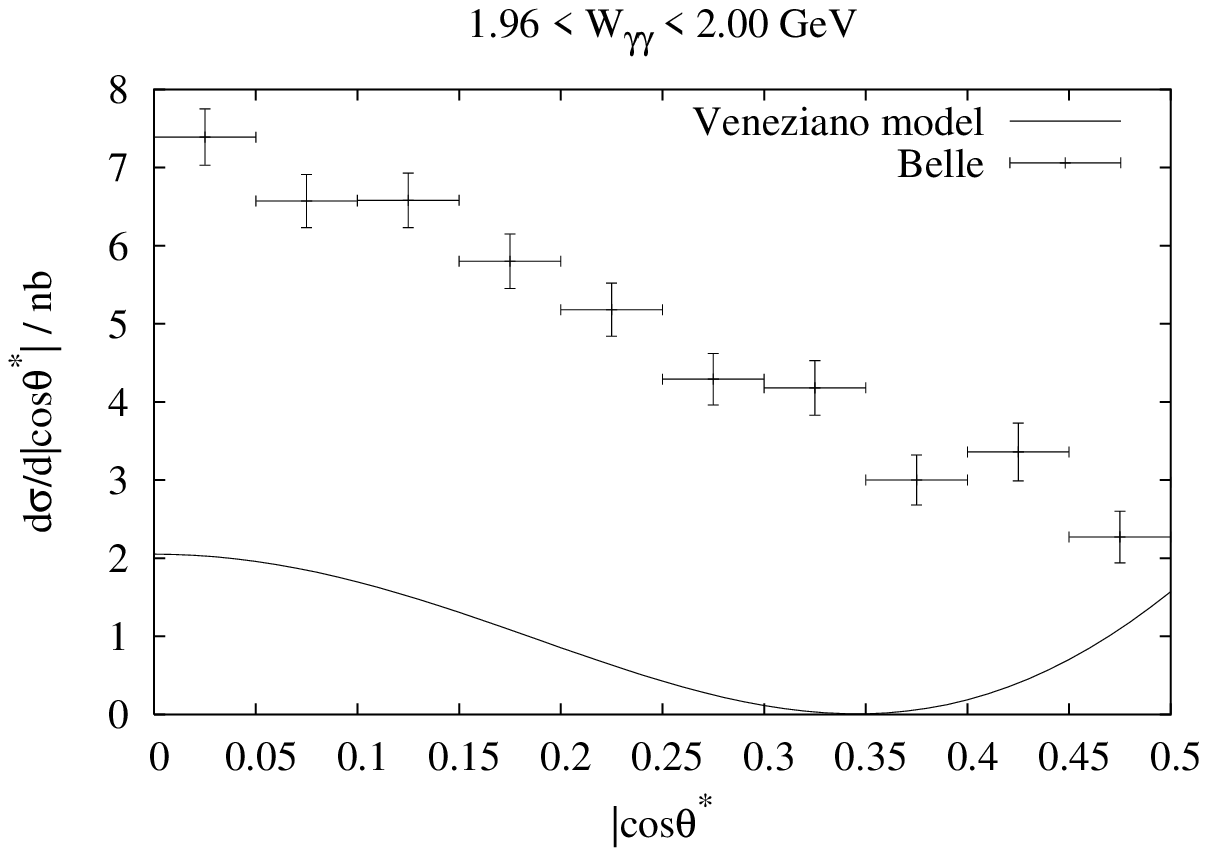,width=3.3cm}
 \epsfig{file=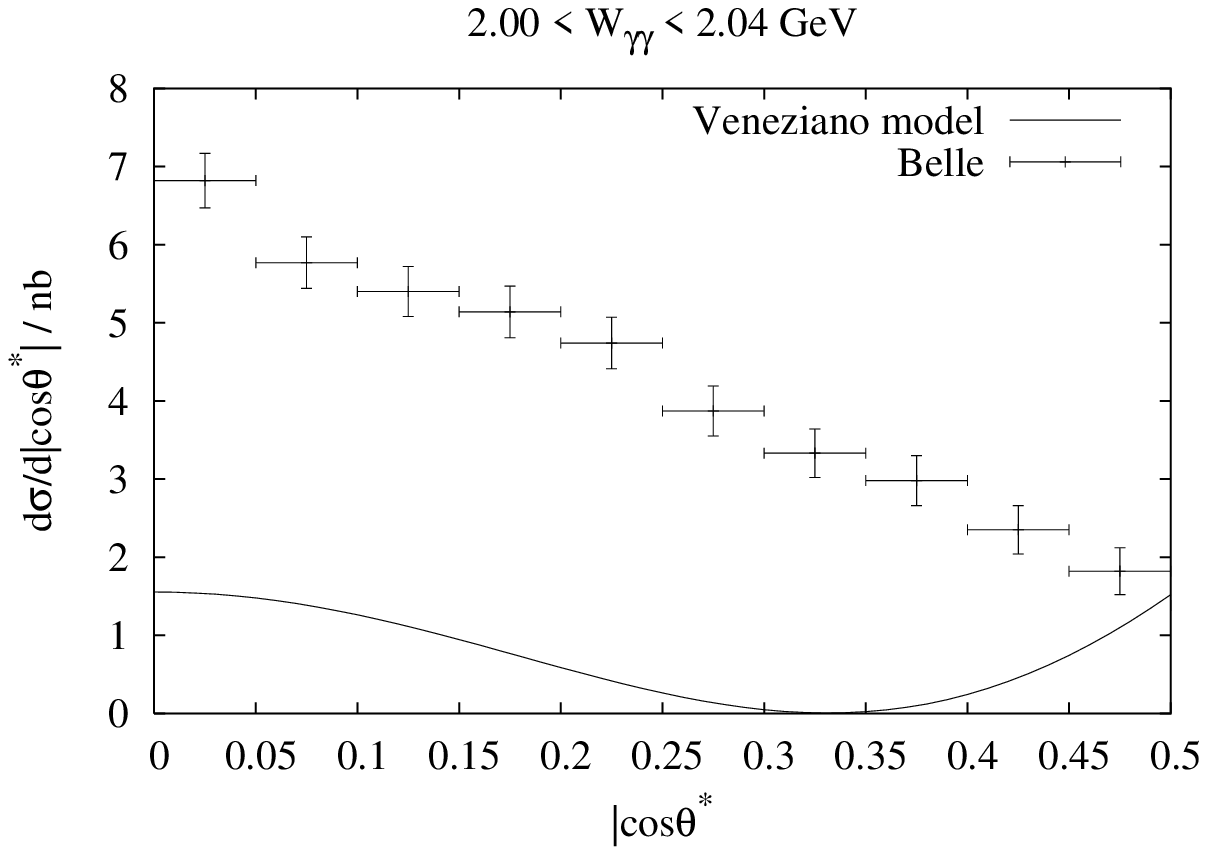,width=3.3cm}

 \epsfig{file=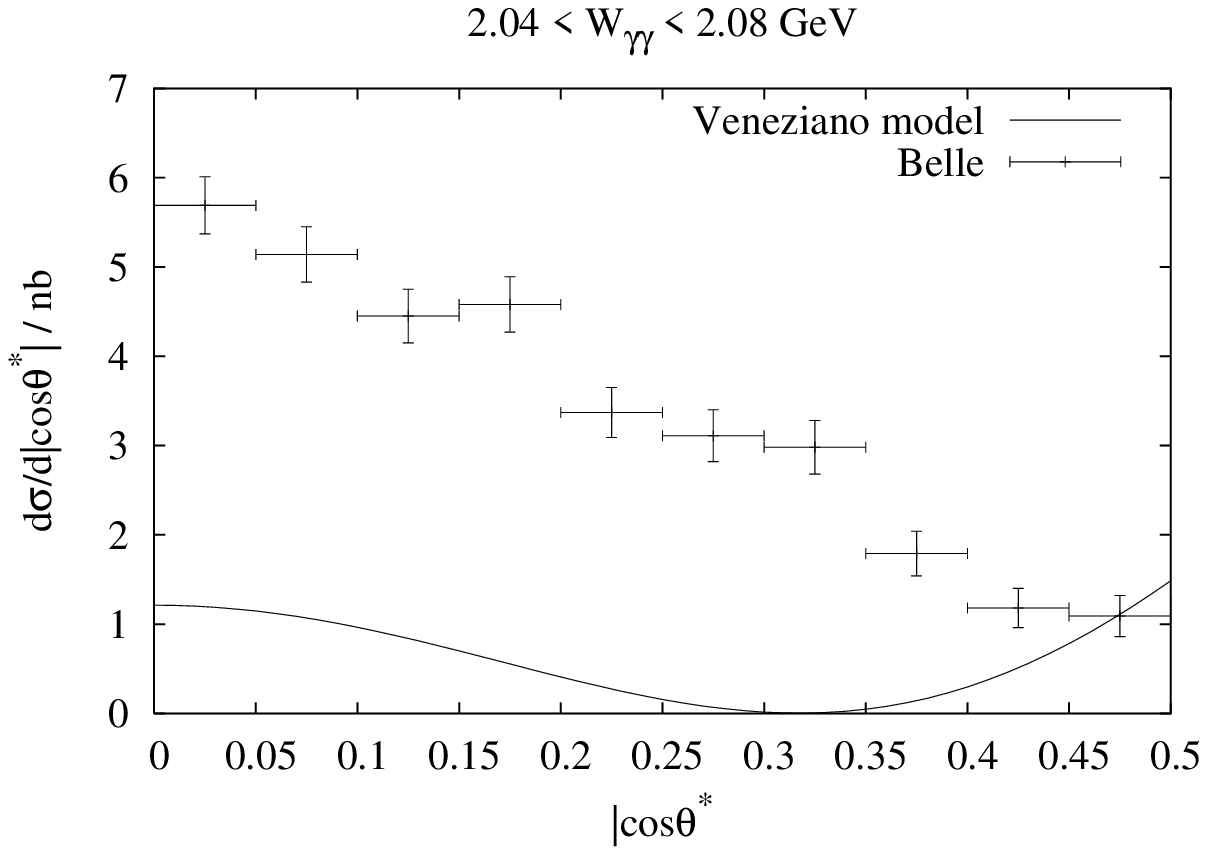,width=3.3cm}
 \epsfig{file=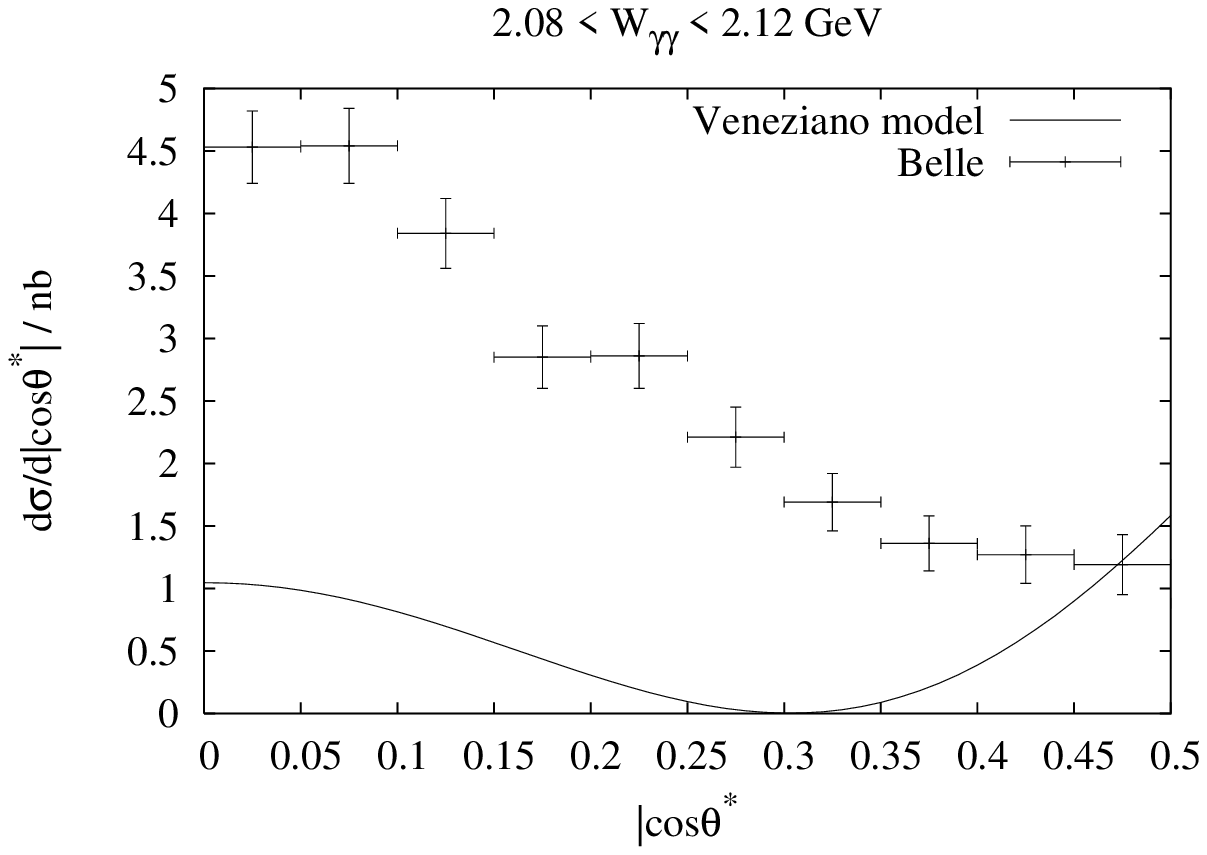,width=3.3cm}
 \epsfig{file=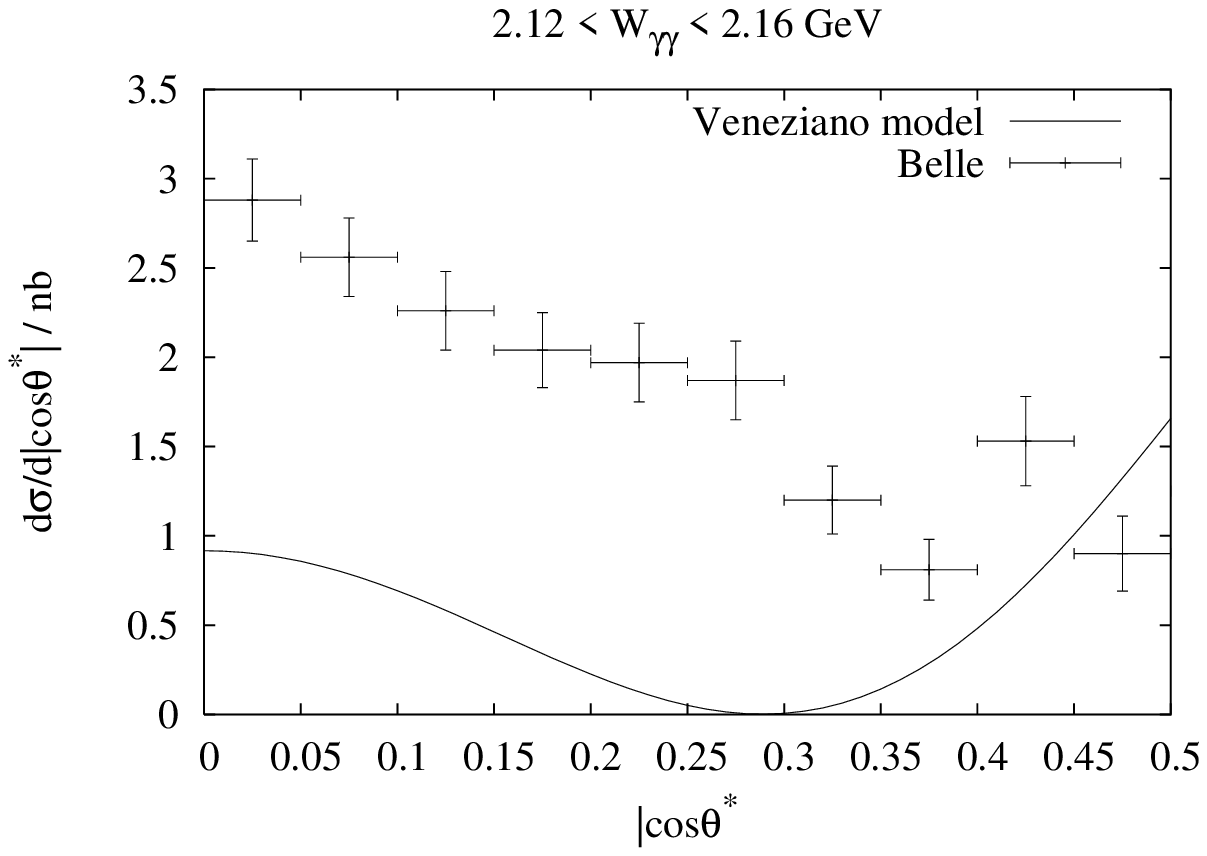,width=3.3cm}
 \epsfig{file=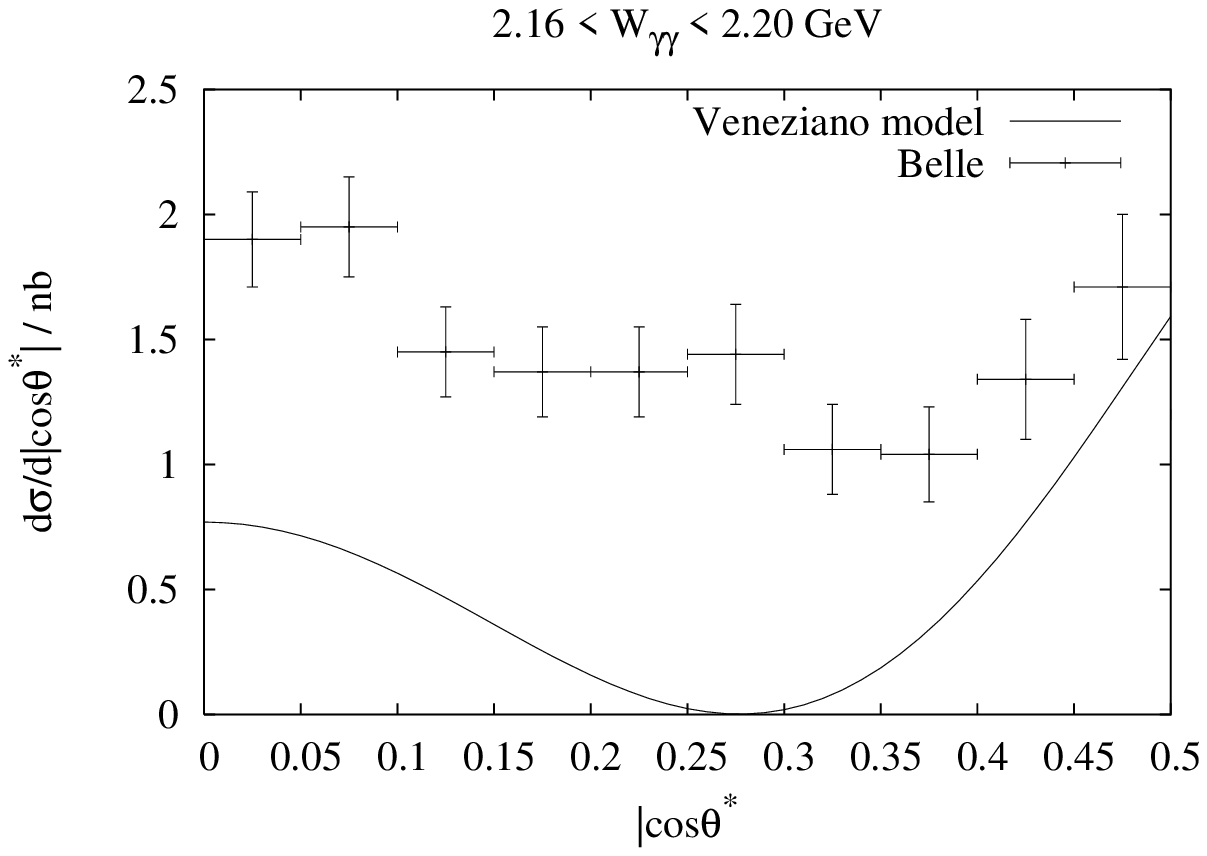,width=3.3cm}

 \epsfig{file=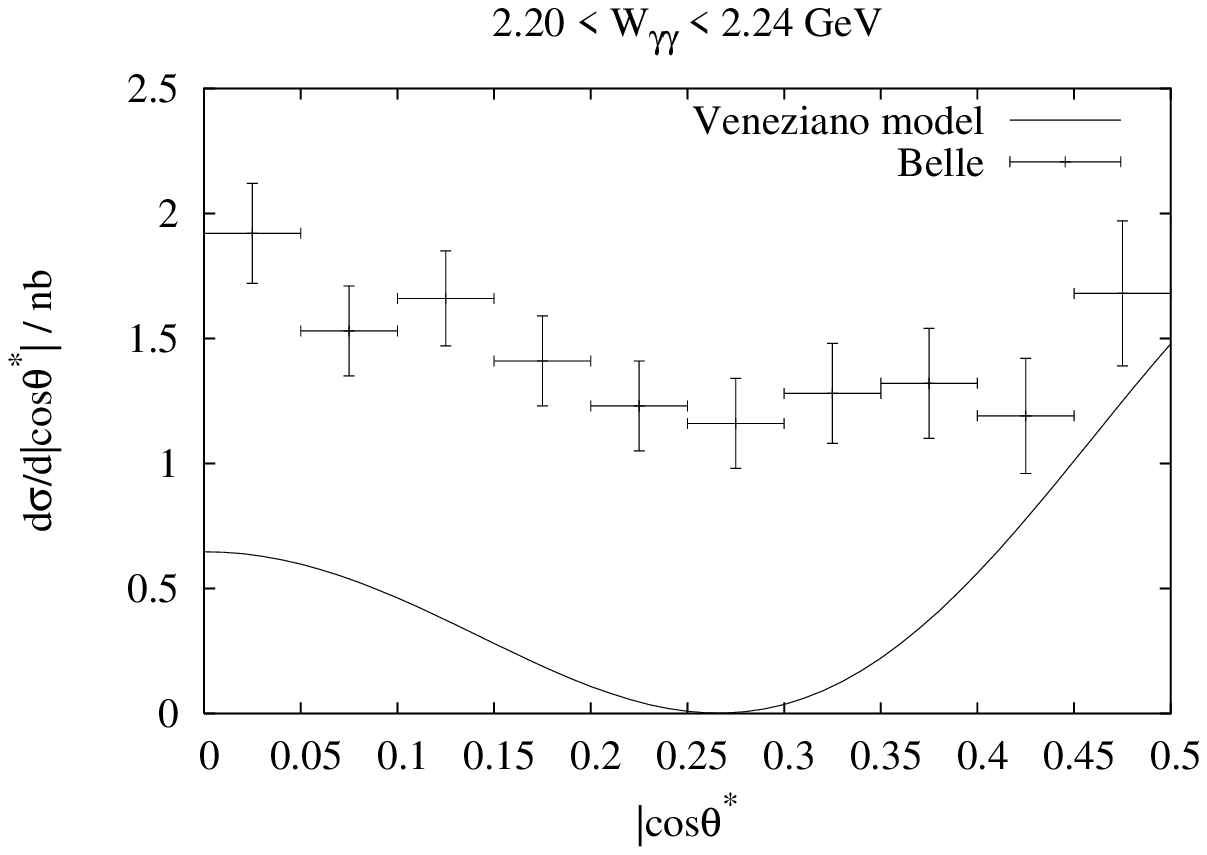,width=3.3cm}
 \epsfig{file=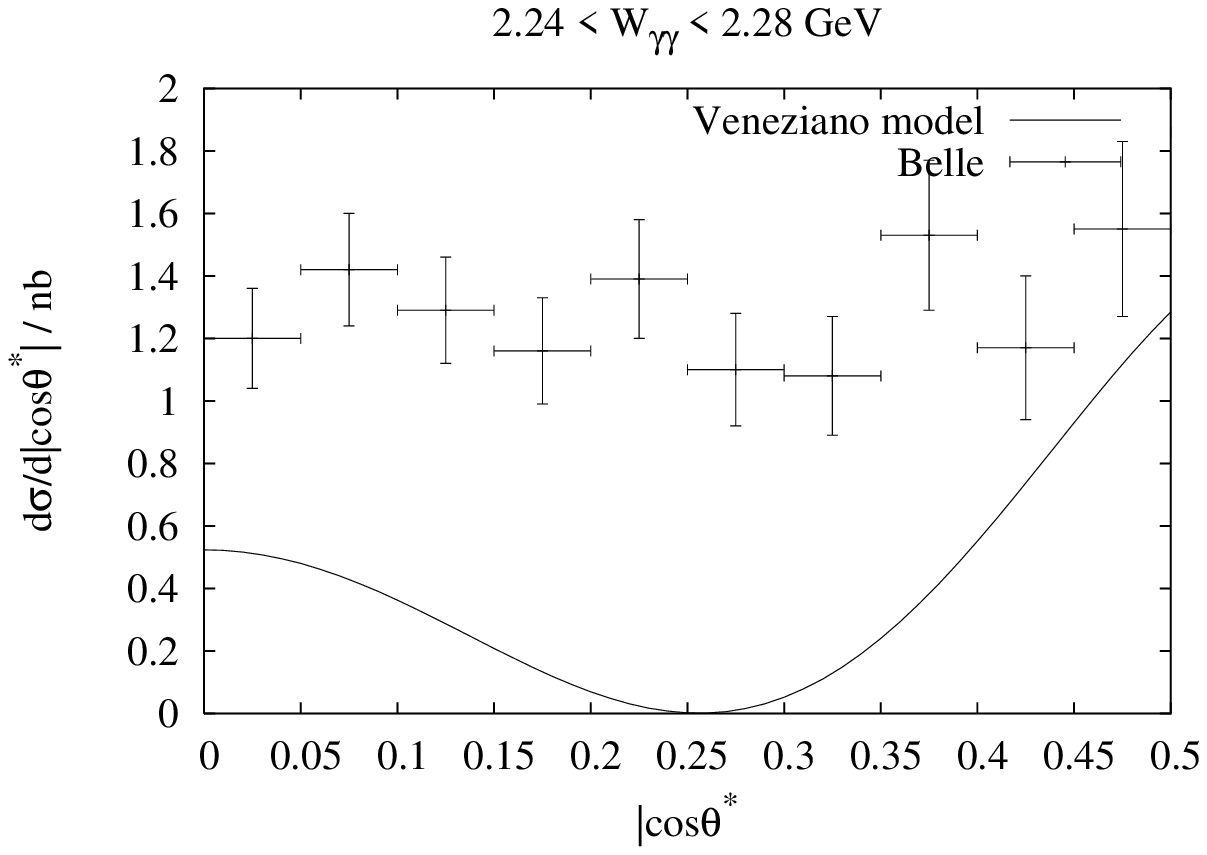,width=3.3cm}
 \epsfig{file=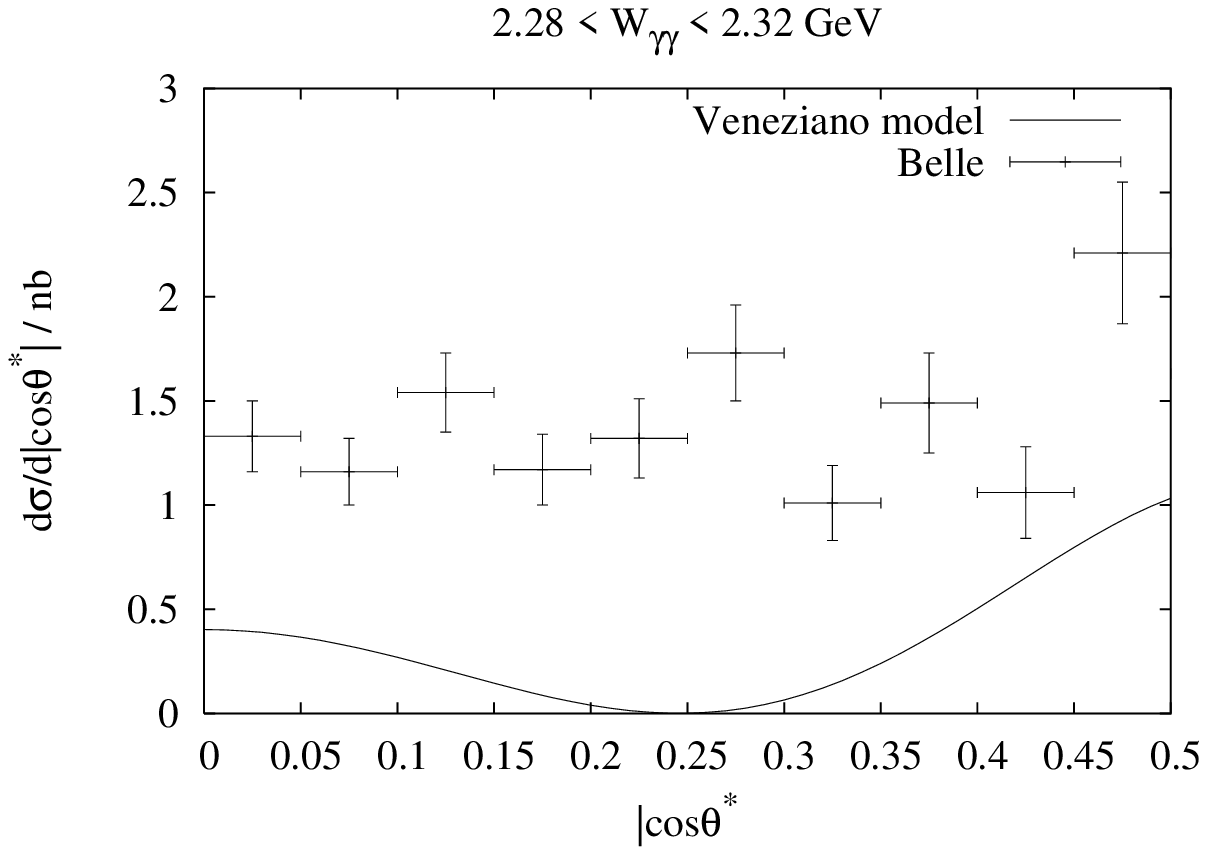,width=3.3cm}
 \epsfig{file=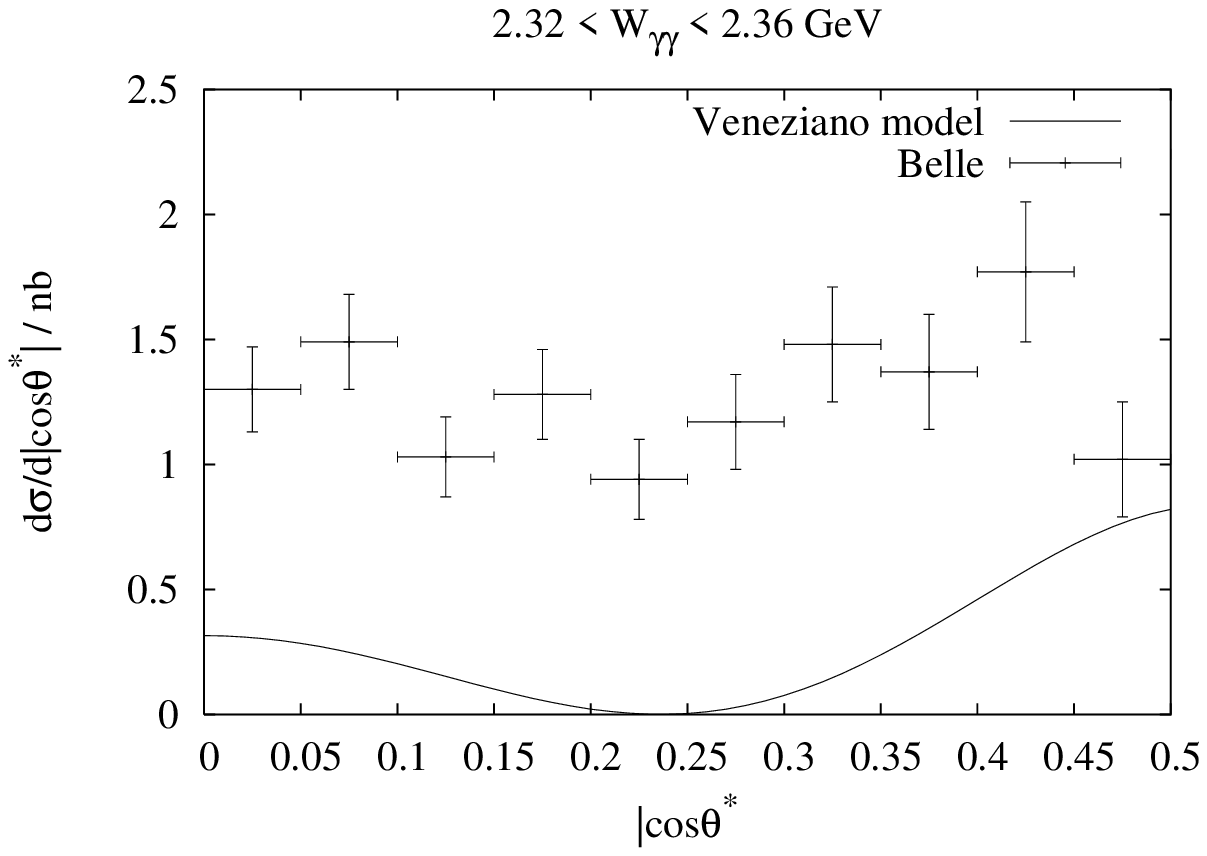,width=3.3cm}

 \epsfig{file=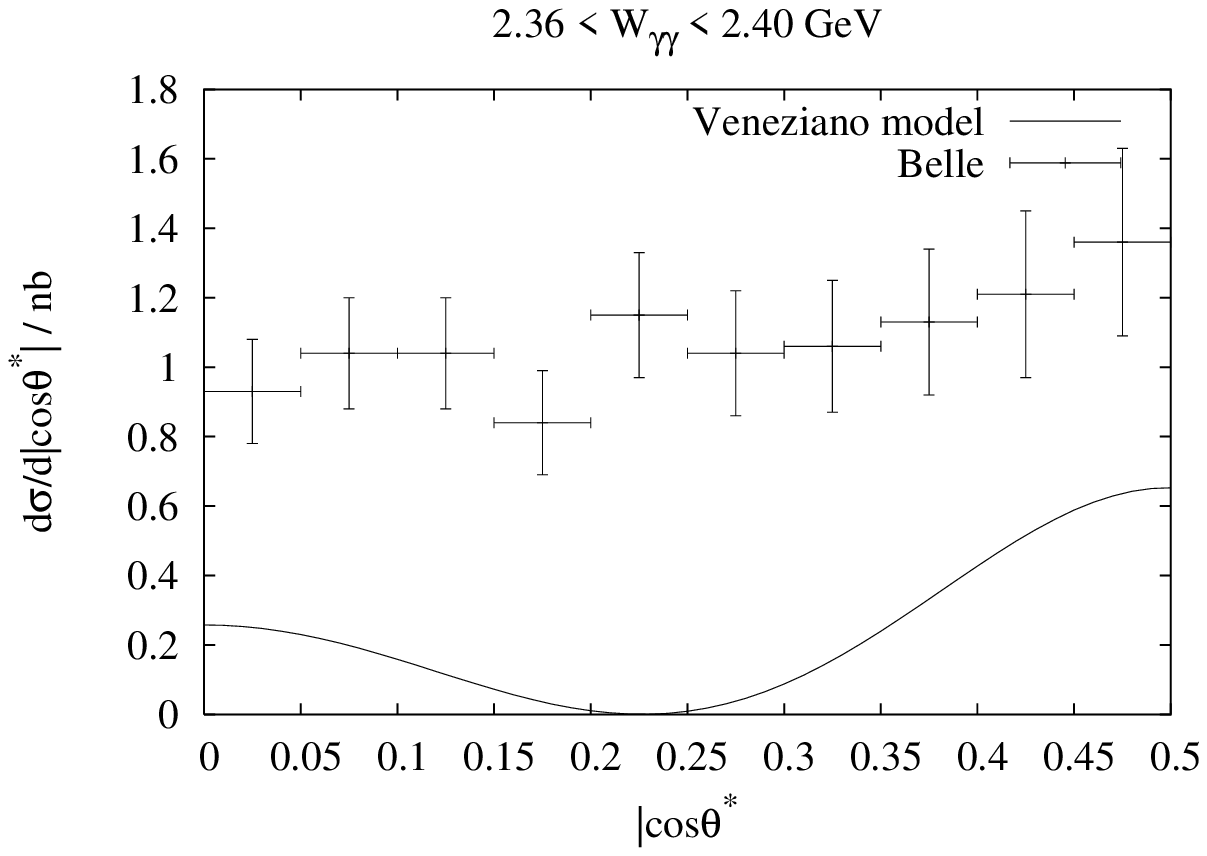,width=3.3cm}
 \caption{The angular distribution for $\gamma\gamma\to K^+K^-$. The 
vertical error-bars on the Belle data are statistical only.
 \label{fig_kpkm_ven_ad}}}

  We show the calculated $\gamma\gamma\to K^+K^-$ cross section in
figs.~\ref{fig_kpkm_ven_cs} and \ref{fig_kpkm_ven_ad}.

  As noted in sec.~\ref{sec_normalization}, we estimate the coupling of 
the $f$ trajectory by means of Regge factorization using the $pK^\pm$ 
scattering total cross sections, and estimate the $f'$ trajectory by 
multiplying the resulting $\overline\beta$ by 1/2.5 to account for the 
charge of the $s$-quark. This last procedure is highly arbitrary, but we 
see that the resulting normalization and the peak positions are roughly in 
agreement with the data.

  There is one unwelcome feature here that the fall-off is more rapid in 
the calculation than real data. One possible reason is the onset of the 
perturbative contribution, but there may be further contribution from 
inaccurate parametrization of the resonances and perhaps extra resonances. 

  The supposition of extra resonances indeed seems quite compelling from 
the richness of mesonic resonances in this region. Otherwise, the 
near-flat structure seen in real data between 1.6 GeV and 2 GeV, and 
between 2.2 GeV and 2.4 GeV, is difficult to explain.

  There is further support to this hypothesis from the angular 
distribution in the sense that we expect perturbative contribution to be 
predominantly in the forward direction, whereas the difference between the 
calculation and the data in reality seems more flat.

  The angular distribution again reproduces the general trend of the 
transition from the resonance picture to the $t$-channel picture. As seen 
before in the $\gamma\gamma\to p\bar p$ case, the $t$-channel behaviour 
seems to set in earlier in the calculation. However, this deficiency is 
less marked in the present case. We note that the baryonic amplitude 
written in eqn.~(\ref{eqn_baryonamp}) does not reproduce the correct Regge 
pole behaviour and has an extra factor of $\sqrt{s}$ in the Regge limit of 
small $t$ and large $s$, whereas the mesonic amplitude of 
eqn.~(\ref{eqn_mesonamp}) has the correct Regge pole behaviour in the 
Regge limit. This may explain the difference.

  If so, an amplitude with the correct Regge behaviour may be able to 
better reproduce the $\gamma\gamma\to p\bar p$ amplitude. We have indeed 
attempted to do this. Our procedure is based on writing several terms of 
Veneziano amplitudes that are proportional to $\left<p|(\not 
p_{\gamma_1}-\not p_{\gamma_2})|\bar p\right>$ and $\left<p|m_p|\bar 
p\right>$. We have attained some success, but we hesitate to reproduce the 
results here as there is too much arbitrariness in this procedure and our 
findings do not show sufficient improvement that compensates for this 
arbitrariness. Indeed, we shall see in sec.~\ref{sec_semilocal} that the 
correct Regge expression by itself is insufficient to account for the 
premature rise of the forward cross section.

  We note again that the dips in the $\cos\theta^*$ distribution are in 
general more pronounced in the calculation than in the real data. However, 
this is mainly for larger values of $W_{\gamma\gamma}$, where the 
perturbative contribution is expected to be important.

 \subsection{$\gamma\gamma\to\Sigma\overline\Sigma,\Lambda\overline\Lambda$}

 \FIGURE[ht]{
 \epsfig{file=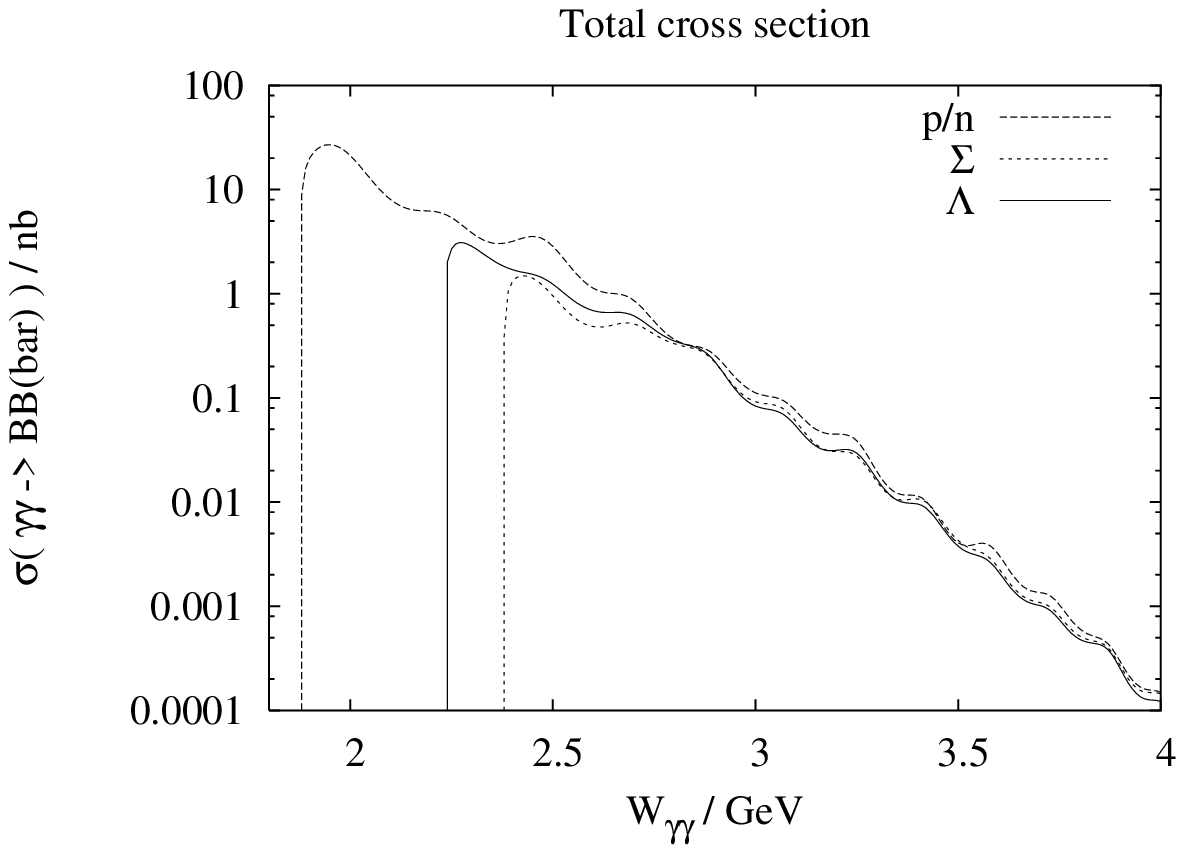,width=14cm}\\
 \epsfig{file=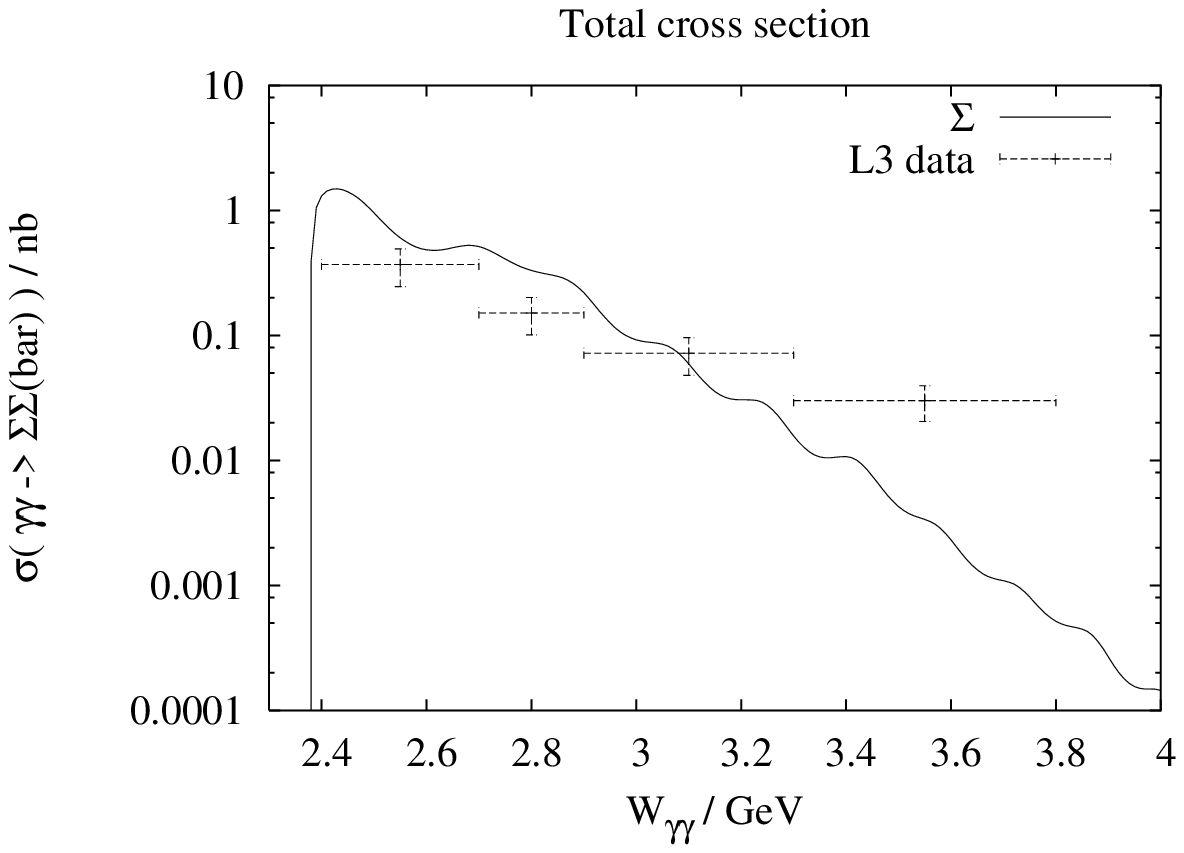,width=7cm}
 \epsfig{file=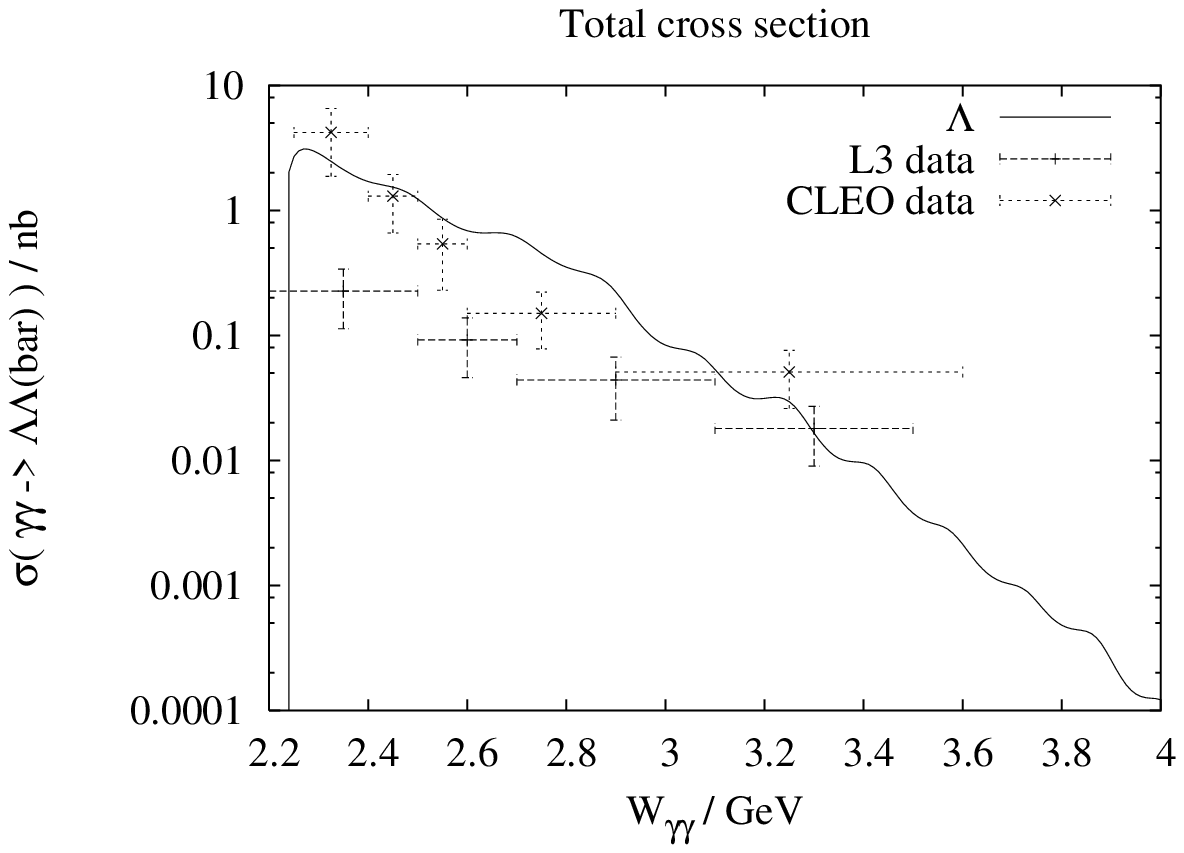,width=7cm}
 \caption{Pair production total cross section, in the region 
$|\cos\theta^*|<0.6$, for $p\bar p$ (top), $\Sigma\overline\Sigma$ (top 
and bottom left) and $\Lambda\overline\Lambda$ (top and bottom right). A 
comparison is made with the data from L3 \cite{baryon_l3} and CLEO 
\cite{baryon_cleo}. The vertical error-bars represent the statistical and 
systematic errors added in quadrature.
 \label{fig_baryon_ven_cs}}}

  We make use of the estimations of sec.~\ref{sec_additive_quark} to 
predict the $\Sigma$ and the $\Lambda$ pair production cross sections in 
fig.~\ref{fig_baryon_ven_cs}. Again, the $f'$ couplings are estimated from 
the charge of the $s$-quark. We expect, as noted in 
sec.~\ref{sec_additive_quark}, that the $\Sigma^\pm$ production cross 
sections are similar to the $\Sigma$ production cross section.

  We should mention that the threshold behaviour is very likely to be 
incorrect in this case as was in the $p\bar p$ case.

  A comparison with the measurements at CLEO \cite{baryon_cleo} of 
$\gamma\gamma\to\Lambda\overline\Lambda$ and L3 \cite{baryon_l3} of 
$\gamma\gamma\to\Sigma^0\overline{\Sigma^0},\Lambda\overline\Lambda$ shows 
that the numbers seem to be larger than the L3 cross sections and more in 
agreement with the CLEO measurement. However, as discussed in 
sec.~\ref{sec_additive_quark}, the normalization for 
$\Lambda\overline\Lambda$ is doubtful because we have not taken into 
account the isospin-0 nature of $\Lambda$. By comparing the normalizations 
of the theory predictions for $\Sigma$ and $\Lambda$ with the 
corresponding data from L3, it seems likely that either the normalization 
for $\Lambda$ production is too high, or the normalization for $\Sigma$ 
production is too low.

  Again, above 3 GeV or so, the description is expected to deteriorate 
because of the effect of perturbative QCD.

 \subsection{The Regge limit and semi-local duality}\label{sec_semilocal}

  One way to understand the above findings is to consider the limiting 
expressions given in eqns.~(\ref{eqn_mesonic_regge}) and 
(\ref{eqn_baryonic_regge}).

  As stated before, photoproduction reactions are not described well in 
the Regge pole picture. However, the use of the Veneziano model is 
motivated by the resonance-pole duality picture, which in turn is 
implied by the finite-energy sum rules \cite{finite_energy_sumrule}.

  Semi-local duality is a generalization of the finite-energy sum rules. 
It states that fluctuations due to resonances cancel out when integrated 
over a finite energy range and the average is given by the Regge 
expression, such as eqn.~(\ref{eqn_mesonic_regge}).

  A comparison of eqn.~(\ref{eqn_baryonic_regge}) with data shows that 
there is in fact not much difference between the result of this expression 
and the result of the full Veneziano expression given by 
eqn.~(\ref{eqn_baryonamp}). However, all the resonances are smoothed out 
so that the curve is more in agreement with experimental data, and the 
angular distribution for low $W_{\gamma\gamma}$ is more flat contrary to 
the real data.

  The central peak is still visible for low $W_{\gamma\gamma}$. This is 
interesting, since Regge expression is merely a sum of $u$- and 
$t$-channel terms. As we have mentioned before, the central peak is best 
understood to be due to the $s$-channel resonance behaviour. This is 
presumably best understood to be an outcome of duality. The sum of 
$t$-channel poles is in fact equivalent to the sum of $s$-channel 
resonance terms.

  It is tempting to compare the data against the correct Regge expression, 
namely:
 \begin{equation}
  A(\gamma\gamma\to B\overline B)=\frac{\overline\beta}\pi
  \Gamma\left(\frac12-\alpha_t(t)\right)
  (-\alpha_s(s))^{\alpha_t(t)}
  + (t\leftrightarrow u).
  \label{eqn_correct_baryonic_regge}
 \end{equation}
  The form of the equation is similar to eqn.~(\ref{eqn_mesonic_regge}), 
and differs from eqn.~(\ref{eqn_baryonic_regge}) by a factor which is 
roughly $\sqrt{-s}$.
  The normalization is different from before, and $\overline\beta$ must be 
determined by other means, such as by fitting with the data. We find that 
$\overline\beta$ needs to be approximately doubled in order to achieve 
correct normalization.

 \FIGURE[ht]{
 \epsfig{file=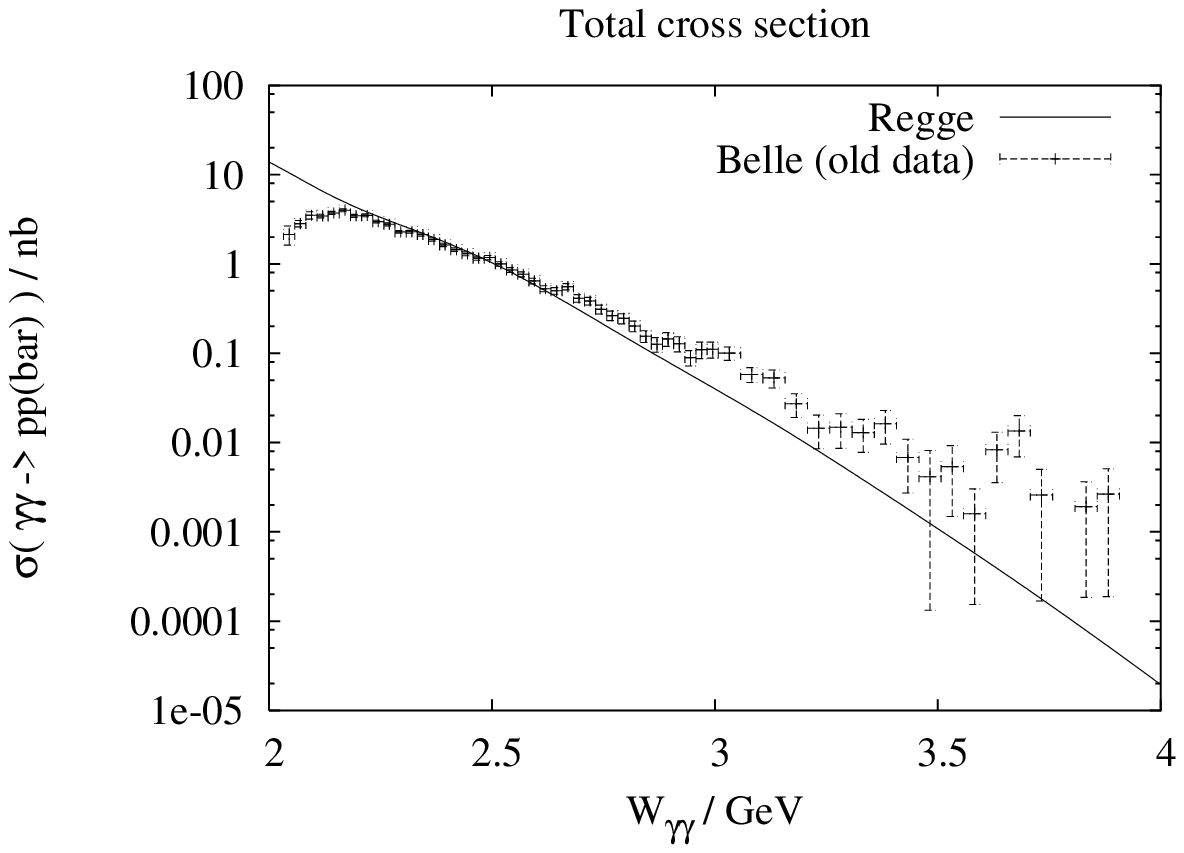,width=14cm}
 \caption{The $\gamma\gamma\to p\bar p$ total cross section in the region
$|\cos\theta^*|<0.6$, calculated using the Regge limiting expression.
  The Belle result is shown for comparison. The vertical error-bars 
represent the statistical uncertainty only.
 \label{fig_regge_cs}}}

  We have calculated the result of this expression and find again that 
there is not much difference between this and the other two expressions. 
  The total cross section is shown in fig.~\ref{fig_regge_cs}, and is in 
good agreement with the data above the threshold region and below about 3 
GeV.
  The angular distribution is in better agreement with the data than that 
calculated using the wrong Regge limit of eqn.~(\ref{eqn_baryonic_regge}). 
The secondary dips that are present above 2.8 GeV in fig.~\ref{fig_ven_ad} 
are not seen in this case.

  What these observations seem to imply is that the wrong Veneziano 
expression of eqn.~(\ref{eqn_baryonamp}) nevertheless works, because the 
amplitude looks like the correct baryonic amplitude whose limiting 
behaviour is given by the Regge expression of 
eqn.~(\ref{eqn_correct_baryonic_regge}).

  Even so, the question remains as to why the Regge approach works in this 
case in the central region, when it is known that photoproduction and 
baryon-exchange reactions are not described well by the Regge pole 
picture.

 \section{Conclusions}\label{sec_conclusions}

  We have shown that an approach based on the resonance-and-pole picture, 
making use of the Veneziano model, provides a sound description of the 
exclusive hadron-pair photoproduction processes both for mesons and 
baryons.

  We have concentrated on the case of $\gamma\gamma\to p\bar p$ and
$\gamma\gamma\to K^+K^-$ at Belle.

  For $\gamma\gamma\to p\bar p$, there is no adjustable parameter after 
normalizing the coupling constant against the $\gamma p$ total cross 
section. For $\gamma\gamma\to K^+K^-$, we obtained the coupling constant 
using Regge factorization and estimated the contribution of the $f'$ 
resonances by the charge-counting argument. We also provided rough 
estimates based on the additive quark rule for 
$\gamma\gamma\to\Sigma\overline\Sigma$ and 
$\gamma\gamma\to\Lambda\overline\Lambda$. For the last of these, we noted 
that there is a problem due to the isospin-0 nature of $\Lambda$.

  Based on the suppression of the coupling of the $a$ trajectory to the 
proton, which follows from the similarity of the $pp,\bar pp$ and the 
$pn,\bar pn$ cross sections, we predict that the $\gamma\gamma\to n\bar n$ 
cross section is similar to the $p\bar p$ cross section, and that there 
would be little difference between $\Sigma^-,\Sigma^0$ and $\Sigma^+$.

  A notable attraction of this procedure is that the transition from the 
resonance picture with a central peak to the $t$-channel forward dominance 
follows naturally.
  On the other hand, we know that the simple Regge pole picture, inherent 
in the Veneziano model, fails for photoproduction, and possibly as a 
result, the $t$-channel behaviour seems to set in earlier than in the real 
data. This is particularly so for the case of $\gamma\gamma\to p\bar p$ 
for which the Veneziano amplitude does not have the correct Regge pole 
behaviour.

  We have made a comparison against the correct limiting expression as 
expected in the Regge picture, and found that the resulting distributions 
are similar to that obtained using the full Veneziano amplitude except 
that the resonance peaks are smoothed out. This similarity of the 
Veneziano amplitude and the correct Regge limiting expression provides 
partial answer as to why our method, which is based on a formula that does 
not have the correct Regge limit, nevertheless works. However, it is not 
clear why the expression for the Regge limit works, when it is known that 
in photoproduction, the Regge pole picture fails in the Regge limit.

  The dips in the angular distribution are more pronounced compared with 
the real data. We have suggested a possible reason for this, based on the 
contribution of different helicity components.

  When the transverse momentum is large, perturbative QCD is expected to 
become the more dominant mode of interaction. We do not have a 
prescription to incorporate this contribution.

  We finally note that final states with more than two hadrons can also be 
estimated in this framework simply by adopting the multi-meson 
generalizations \cite{koba-nielsen} of the Veneziano model. This should be 
applicable again in the central region before the onset of perturbative 
QCD. This would induce one free parameter, namely the normalization, which 
we presently do not know how to estimate.

 \acknowledgments

  We thank Augustine Chen, Chuan-Hung Chen, Wan-Ting Chen, Chen-Cheng
Kuo and Yoshiaki Yasui for discussions.

 \end{document}